\documentclass[useAMS,usenatbib]{mn2e}

\usepackage[T1]{fontenc}
\usepackage{ae,aecompl}
\usepackage{times}

\usepackage{graphicx}
\usepackage{dcolumn}
\usepackage{bm}
\usepackage{listings}
\usepackage{amssymb}
\usepackage{verbatim}
\usepackage{url}
\usepackage{mathtools}
\usepackage{amsmath}
\usepackage{longtable}
\usepackage{booktabs}
\usepackage{ctable}
\usepackage{natbib}
\usepackage{hyperref}

\makeatletter
\newcommand{\HI}{{\rm H\,{\scriptstyle I}}}
\newcommand{\HII}{{\rm H\,{\scriptstyle II}}}
\newcommand{\HeI}{{\rm He\,{\scriptstyle I}}}
\newcommand{\HeII}{{\rm He\,{\scriptstyle II}}}
\newcommand{\HeIII}{{\rm He\,{\scriptstyle III}}}
\newcommand{\rmnum}[1]{\romannumeral #1}
\newcommand{\Rmnum}[1]{\expandafter\@slowromancap\romannumeral #1@}
\newcommand{\nH}{n_{\mbox{\tiny H}}}

\newcommand{\xHII}{x_{\mbox{\tiny $\HII$}}}
\newcommand{\xHeII}{x_{\mbox{\tiny $\HeII$}}}
\newcommand{\xHeIII}{x_{\mbox{\tiny $\HeIII$}}}

\newcommand{\LyA}{\mbox{Ly}\alpha}

\newcommand{\NHI}{N_{\mbox{\tiny H\Rmnum{1}}}}

\makeatother

\title[Ionization and temperature around high-z QSOs]
  {The Concerted Impact of Galaxies and QSOs on the Ionization and Thermal State of the Intergalactic Medium}

\author[K. Kakiichi et al.]
{Koki Kakiichi$^{1}$\thanks{E-mail: kakiichi@mpa-garching.mpg.de},
Luca Graziani$^{1,2}$, 
Benedetta Ciardi$^{1}$,
Avery Meiksin$^{3}$, 
\newauthor Michele Compostella$^{1}$, 
Marius B. Eide$^{1}$,
Saleem Zaroubi$^{4}$
\\
$^{1}$ Max Planck Institute for Astrophysics, Karl-Schwarzschild Stra\ss e 1, 85741 Garching, Germany\\
$^{2}$ INAF Osservatorio Astronomico di Roma, Via Frascati 33, 00040, Monte Porzio Catone (RM), Italy\\
$^{3}$ SUPA, Institute for Astronomy, University of Edinburgh, Blackford Hill, Edinburgh\, EH9\ 3HJ, UK \\
$^{4}$ Kapteyn Astronomical Institute, University of Groningen, PO Box 800, NL-9700 AV Groningen, the Netherlands\\
}

\date{Released 2016 Xxxxx XX}

\pagerange{\pageref{firstpage}--\pageref{lastpage}} \pubyear{2015}

\def\LaTeX{L\kern-.36em\raise.3ex\hbox{a}\kern-.15em
    T\kern-.1667em\lower.7ex\hbox{E}\kern-.125emX}

\begin{document}

\maketitle

\begin{abstract}
We present a detailed analysis of the ionization and thermal structure of the intergalactic medium (IGM) around a high-redshift QSO using a large suite of cosmological, multi-frequency radiative transfer (RT) simulations, exploring the contribution from galaxies as well as the QSO, and the effect of X-rays and secondary ionization. We show that in high-$z$ QSO environments both the central QSO and the surrounding galaxies concertedly control the reionization morphology of hydrogen and helium and have a non-linear impact on the thermal structure of the IGM. A QSO imprints a distinctive morphology on $\HII$ regions if its total ionizing photon budget exceeds that of the surrounding galaxies since the onset of hydrogen reionization; otherwise, the morphology shows little difference from that of $\HII$ regions produced only by galaxies. In addition, the spectral shape of the collective radiation field from galaxies and QSOs controls the thickness of the I-fronts. While a UV-obscured QSO can broaden the I-front, the contribution from other UV sources, either galaxies or unobscured QSO, is sufficient to maintain a sharp I-front. X-rays photons from the QSO are responsible for a prominent extended tail of partial ionization ahead of the I-front. QSOs leave a unique imprint on the morphology of $\HeII/\HeIII$ regions. We suggest that, while the physical state of the IGM is modified by QSOs, the most direct test to understand the role of galaxies and QSOs during reionization is to perform galaxy surveys in a region of sky imaged by 21 cm tomography. 
\end{abstract}


\section{Introduction}

In the current paradigm of extragalactic astronomy, the Universe has undergone two major reionization epochs. $\HI$ and $\HeI$ reionization are thought to occur at $z \sim 6-20$, driven by early star forming galaxies and/or quasars (QSOs) (e.g. \citealt{2015ApJ...802L..19R,2015ApJ...813L...8M}). This first reionization epoch contains information about the formation of the first luminous objects in the Universe, a period called Cosmic Dawn. The second reionization epoch, from $\HeII$ to $\HeIII$, occurs later time, at $z\sim2-4$, and it is likely driven by the high quasar activity near the peak of cosmic star formation history. Understanding the full reionization process of intergalactic hydrogen and helium will provide a milestone in investigating the nature of high-redshift galaxies and QSOs and of their interaction with the intergalactic medium (IGM).

Current observations of the $\HI$ $\LyA$ Gunn-Peterson trough suggest that hydrogen reionization is largely completed by $z \sim 6$ (e.g. \citealt{2006AJ....132..117F}). However, the nature of the sources that drive hydrogen reionization is still unknown. A concordance picture is one in which (undetected) faint star forming galaxies with an escape fraction of ionizing photons as large as $\sim20$ per cent  are the main reionizing agents (e.g. \citealt{2015ApJ...802L..19R}). At lower redshift, observations of the $\HeII$ $\LyA$ Gunn-Peterson trough suggest that $\HeII$ reionization is mostly completed by $z\sim2.5$ (\citealt{2011ApJ...733L..24W,2011ApJ...726..111S}). Unlike $\HI$ reionization, $\HeII$ reionization is likely driven by the already detected population of QSOs (\citealt{2015A&A...578A..83G}). In fact, simulations calibrated with the observed QSO luminosity function can grossly reproduce the observed properties of the $\HeII$ $\LyA$ forest (\citealt{2009ApJ...694..842M,2013MNRAS.435.3169C,2014MNRAS.445.4186C}). 

As hydrogen and helium reionization have a strong impact on the thermal state of the IGM (\citealt{2006ApJ...644...61L,2011MNRAS.415..977M,2015MNRAS.447.2503G}), constraints on the reionization process can be obtained by measurements of the IGM temperature. Such measurements from the $\LyA$ forest (\citealt{2011MNRAS.415..977M,2014ApJ...788..175L}) suggest that the end of $\HI$ reionization should be at a redshift lower than $z\simeq9$ (\citealt{2002ApJ...567L.103T,2010MNRAS.406..612B,2012MNRAS.421.1969R}), while $\HeII$ reionization should be extended over $2\leq z\leq4.8$ (\citealt{2011MNRAS.410.1096B}). Furthermore, the IGM temperature in QSO near zones at $z\sim6$ suggests that the onset of $\HeII$ reionization may be as early as the formation of the first QSOs at $z>6$ (\citealt{2012MNRAS.419.2880B}). 

Although many authors have studied the impact of QSOs on the ionization and thermal state of the IGM (e.g. \citealt{1987ApJ...321L.107S,1993ApJ...412...34M,1999ApJ...514..648M,2000ApJ...542L..75C,2000ApJ...530....1M,2003ApJ...586..693W,2004ApJ...604..484M,2005ApJ...620...31Y,2005ApJ...623..683Y,2006ApJ...648..922S,2007ApJ...657...15K,2007MNRAS.376L..34M,2007MNRAS.380L..30A,2007MNRAS.380.1369T,2007ApJ...670...39L,2008MNRAS.384.1080T,2008MNRAS.385.1561K,2008ApJ...686...25F,2012MNRAS.424..762D,2013MNRAS.429.1554F,2016MNRAS.455.2778F,2015MNRAS.454..681K}), the role of QSOs during $\HI$ reionization at $z>6$ still remains unclear. While it has been argued that QSOs alone cannot be responsible for hydrogen reionization based on constraints from the unresolved X-ray background (\citealt{2004ApJ...613..646D,2005MNRAS.362L..50S,2007MNRAS.374..761S,2015A&A...575L..16H}) and the decreasing number density of high-$z$ QSOs (\citealt{2005MNRAS.356..596M}), recent re-investigations suggest a possible important contribution (e.g. \citealt{2011ApJ...728L..26G,2012MNRAS.425.1413F,2015A&A...578A..83G,2015ApJ...813L...8M}). 

Furthermore, theoretical predictions about how the QSOs impact on the ionization and thermal state of the local environment do not always agree. For example, there seems to be no consensus about whether a QSO imprints a distinctive ionization and thermal structure on the IGM by producing a very large ionized region (e.g. \citealt{2013MNRAS.429.1554F,2007MNRAS.375.1269Z,2008MNRAS.384.1080T}), if it can be distinguished from one produced only by galaxies (\citealt{2007MNRAS.380L..30A,2007ApJ...670...39L,2012MNRAS.424..762D}), if the spectra of galaxies and QSOs induce a different shape of the ionization front (I-front) (\citealt{2005MNRAS.360L..64Z,2008MNRAS.385.1561K}), or whether the sphericity of an ionized region may serve as discriminator of the type of source that created it (\citealt{2012MNRAS.424..762D}). These theoretical discrepancies must be resolved to interpret QSO absorption spectra and upcoming 21 cm observations. In other words, a correct theoretical understanding is vital to place constraints on the role of galaxies and QSOs in driving reionization.

Understanding the impact of ionization from galaxies and/or QSOs on the physical state of the IGM relies on an accurate modeling of the effects of (\rmnum{1}) the spectral shape of QSOs and galaxies (i.e. multi-frequency radiative transfer [RT]), (\rmnum{2}) galaxies surrounding a QSO  (i.e. cosmological $N$-body/hydrodynamical simulations), (\rmnum{3}) an anisotropic propagation of the I-front (i.e. 3D simulation), and (\rmnum{4}) a coherent treatment of both UV and X-ray photons as well as secondary ionization (i.e. UV/X-ray physics). 
Previous works addressed the above aspects separately. For example,  the 1D RT simulations by \cite{2008MNRAS.384.1080T} and \cite{2008MNRAS.385.1561K} focused on (\rmnum{1}) and (\rmnum{4}), whereas the 3D RT simulation by \cite{2012MNRAS.424..762D} including both galaxies and QSOs, but no thermal structure, focused on addressing (\rmnum{1}), (\rmnum{2}) and (\rmnum{3})\footnote{\cite{2013MNRAS.429.1554F} and \cite{2015MNRAS.454..681K} have also conducted RT simulation post-processing cosmological hydrodynamical simulation, but without simultaneously accounting for both galaxies and QSOs.}. At the time of writing, no radiative transfer calculation addressing all four points in a single simulation is reported.

In this work we therefore investigate the ionization state of both hydrogen and helium, and the thermal state of the IGM in the environment of a high-$z$ QSO by performing a suite of multi-dimensional, multi-frequency radiative transfer simulations post-processing a cosmological hydrodynamical simulation. This is to our knowledge the most detailed calculation of this kind to date. This work distinguishes itself from previous investigations, as we present the case-by-case analysis of hydrogen and helium reionization, as well as the thermal state of the environment of a QSO at $z=10$, including the effect of surrounding galaxies, X-ray and secondary ionization. This large suite of RT simulations aims at providing insights into the underlying physical mechanisms responsible for controlling the physical state of the IGM around QSOs, and on how all processes collectively shape it. Our work is an important first step towards full cosmological reionization simulations including both galaxies and multiple QSOs.

The paper is organized as follows. First we describe the simulation setup in \S~\ref{sec:sim}. In \S~\ref{sec:H_reion}, we present the results of hydrogen and helium reionization. In \S~\ref{sec:thermal}, we discuss the thermal state in the QSO environment. We compare our results with previous works in \S~\ref{sec:comparison}. Observational implications are briefly discussed in \S~\ref{sec:obs}. Conclusions are then presented in \S~\ref{sec:conclusion}.

\section{Simulations}\label{sec:sim}

In the following we will describe the simulations employed in this paper. Hereafter we assume cosmological parameters: $\Omega_\Lambda=0.74$, $\Omega_m=0.26$, $\Omega_b=0.0463$, $h=0.72$, $n_s=0.95$ and $\sigma_8=0.85$, which is consistent with the WMAP9 result (\citealt{2013ApJS..208...19H}).

\subsection{Hydrodynamical simulations}

We have used hydrodynamical simulations of the IGM with the smoothed particle hydrodynamics code GADGET-3, which is an updated version of the publicly available code GADGET-2 (\citealt{2005MNRAS.364.1105S}). The size of the simulation box is $(50h^{-1}\rm~cMpc)^3$, with $2\times512^3$ dark matter and gas particles, corresponding to a mass of $5.53\times 10^7h^{-1}\rm M_\odot$ and $1.20 \times 10^7h^{-1}\rm M_\odot$, respectively. 
The simulation is by design centered on the most massive halo (with mass $1.34\times10^{10}h^{-1}\rm M_\odot$) at $z=10$.
Haloes are identified using the friend-of-friends algorithm with a linking length of 0.2 times the mean inter-particle separation. 
All haloes are considered to harbor one galaxy with the empirically motivated prescription described in \S~\ref{galaxy_source}, which results in 7534 galaxies at $z=10$. The lowest halo mass at this redshift is $3.9\times10^8h^{-1}\rm M_\odot$.
 
The gas density and temperature are mapped onto Cartesian grids with $256^3$ cells; in total 11 snapshots are present in the redshift range $z=15-10$, with a $\Delta z=0.5$ interval. These are used as input for the RT code \texttt{CRASH} (see \S~\ref{sec:RT}).

\subsection{Multi-frequency radiative transfer simulations}\label{sec:RT}

The outputs of the hydro simulations described above are post-processed with the Monte Carlo RT code \texttt{CRASH} (\citealt{2001MNRAS.324..381C,2003MNRAS.345..379M,2009MNRAS.393..171M,2013MNRAS.431..722G}; Graziani et al. in prep), which, in its latest version, solves the cosmological radiative transfer of UV and soft X-ray photons\footnote{We refer to photons in the energy range 13.6~eV-200~eV (200~eV-2~keV) as UV (soft X-ray) photons.} in a gas made of hydrogen and helium. We refer the reader to the original papers for more details on the code. Here we assume a H and He number fraction of $X=0.92$ and $Y=0.08$, respectively. 

The code self-consistently evolves the ionization state of hydrogen and helium and recalculates the gas temperature by solving for non-equilibrium chemistry and radiative heating and cooling processes. The photoionization heating of $\HI$, $\HeI$, and $\HeII$, recombination cooling, and adiabatic cooling by cosmological expansion are self-consistently included within the RT calculation (see \citealt{2003MNRAS.345..379M,2013MNRAS.431..722G}; Graziani et al. in prep for a more comprehensive list of heating and cooling mechanisms implemented in the code). 

For our multi-frequency RT calculations, the source spectrum from 13.6~eV to 2~keV is sampled by 42 frequency bins using 29 bins for the UV range, with finer sampling near the ionization threshold of $\HI$ (13.6 eV), $\HeI$ (24.6 eV) and $\HeII$ (54.4 eV), and 13 bins to cover the soft X-ray range. We have tested the sensitivity of the results to the upper frequency adopted, using $1,~2$ and $4\rm~keV$. We find that the ionization and thermal structure within $\sim 20h^{-1}\rm cMpc$ is insensitive to the upper cut, and we have thus chosen $2\rm~keV$. 
The Monte Carlo simulations cast $10^6$ photon packets per galaxy and $2\times10^8$ per QSO. We have verified the convergence of the results in Appendix \ref{app:photon_packet}.  

The radiative transfer of X-ray photons and the effect of fast photo-electrons produced by high energy photons are described in detail in Graziani et al. (2016, in prep.). Secondary ionization effects have been modelled by many authors (e.g. \citealt{1985ApJ...298..268S,1999ApJS..125..237D,2008MNRAS.387L...8V,2010MNRAS.404.1869F}). Although the results are qualitatively similar among different approaches, there are some  quantitative differences in the IGM thermal and ionization states. Here we adopt the model by \citet{2008MNRAS.387L...8V}. We refer the reader to Graziani et al. (2016, in prep.) for a more comprehensive comparative analysis.

\begin{table*}
\caption[Main simulation parameters]{Parameters used for the simulations described in the text. From left to right, the columns refer to the simulation ID; the normalization of the total ionizing photon emissivity of galaxies, $f_{\rm UV}$; the galaxy/QSO spectrum slope $\alpha_{G}$/$\alpha_{Q}$; the inclusion/exclusion of the reionization history from galaxies in the range $z=15-10$ (`history' column); the presence of a QSO; the inclusion of X-ray photons and of secondary ionization. The one highlighted in boldface (GAL\_R+QSO\_UVXsec) is our reference run.}\label{table:2}
    \begin{tabular}{lllllll}
    \hline
    run ID                  & $f_{\rm UV}$ & $\alpha_{G}$, $\alpha_{Q}$ & history & QSO & X-ray & secondary \\
    \hline
    Galaxies only \\
    \hline
    GAL         & 0.1     & 3,   n/a              & off         & off & off  & off \\
    GAL\_0.5R  & 0.05    & 3,   n/a              & on          & off & off  & off \\
    GAL\_R     & 0.1     & 3,   n/a              & on          & off & off  & off \\
    GAL\_2R    & 0.2     & 3,   n/a              & on          & off & off  & off \\
    \hline
    QSO only \\
    \hline
    QSO\_UV           & n/a      & n/a, 1.5               & off           & on  & off  & off \\
    QSO\_UVX          & n/a      & n/a, 1.5               & off           & on  & on   & off \\
    QSO\_UVXsec       & n/a      & n/a, 1.5               & off           & on  & on   & on  \\
    QSO\_obsc\_UVXsec  & n/a      & n/a, 1.5(obscured)     & off           & on  & on   & on  \\
    \hline
    Galaxies+QSO \\
    \hline
    GAL\_R+QSO\_UV            & 0.1        & 3,   1.5               & on         & on  & off & off \\
    GAL\_R+QSO\_UVX           & 0.1        & 3,   1.5               & on         & on  & on  & off \\
    GAL\_0.5R+QSO\_UVXsec     & 0.05       & 3,   1.5               & on         & on  & on  & on  \\
    {\bf GAL\_R+QSO\_UVXsec}  & {\bf 0.1}  & {\bf 3,   1.5}         & {\bf on}   & {\bf on}  & {\bf on}  & {\bf on}  \\
    GAL\_2R+QSO\_UVXsec       & 0.2        & 3,   1.5               & on         & on  & on  & on  \\
    GAL\_R+QSO\_obsc\_UVXsec    & 0.1        & 3,   1.5(obscured)     & on         & on  & on  & on  \\
    \hline
    \end{tabular}
\end{table*}

We create a suite of 14 RT simulations to study the reionization models by galaxies and QSOs (see Table \ref{table:2}). The procedure adopted is the following.

Our multi-frequency RT simulations consist of two steps. First, we perform three runs from $z=15$ until $z=10$ with only galaxy-type sources. 
The runs differ only in the total ionizing photon emissivities of galaxies (parameterized by $f_{\rm UV}$, see \S~\ref{galaxy_source}). Our fiducial run is indicated by `R' (standing for Reference), while the other two (`0.5R' and `2R') have an emissivity which is 0.5$\times$R and 2$\times$R. The source models of galaxies are described in \S~\ref{galaxy_source}.

Second, we further evolve the final snapshot (at $z=10$) for a time corresponding to the lifetime of the QSO, $t_Q$, including only galaxies (`Galaxies only' runs) or both galaxies and a QSO (`Galaxies+QSO' runs).
The QSO is positioned within the most massive halo ($1.34\times10^{10}h^{-1}\rm M_\odot$) at the center of the simulation box (see \S~\ref{qso_source} for the detail of our QSO model).
As a reference, the QSO alone is also switched on in a fully neutral medium (`QSO only' runs).

We then perform simulations with various combinations of RT processes (with/out photoionization by X-rays and contribution of secondary ionization) to study the impact of the QSO in exquisite details. Our reference run (GAL\_R+QSO\_UVXsec) follows UV and X-ray photons (with secondary ionization) emitted by galaxies as well as by the QSO in a medium ionized by the pre-existing galaxy population. The suite of RT simulations is summarized in Table \ref{table:2}. 

\subsection{Source model: galaxies}\label{galaxy_source}

To assign an ionizing photon production rate to a galaxy we follow the method described in \citet{2012MNRAS.423..558C}, and we refer the reader to this paper for more details. Briefly, the comoving ionizing photon emissivity of the entire galaxy population at $z>6$ is modelled based on a pre-assumed global star formation rate density, observations of the ionizing background from the Ly$\alpha$ forest (\citealt{2007MNRAS.382..325B}) and of high-redshift galaxies (\citealt{2013ApJ...773...75O,2015ApJ...803...34B}).  The total ionizing photons are then distributed among all galaxies in the simulation box.

\begin{figure}
  \centering
  \includegraphics[angle=0,width=\columnwidth]{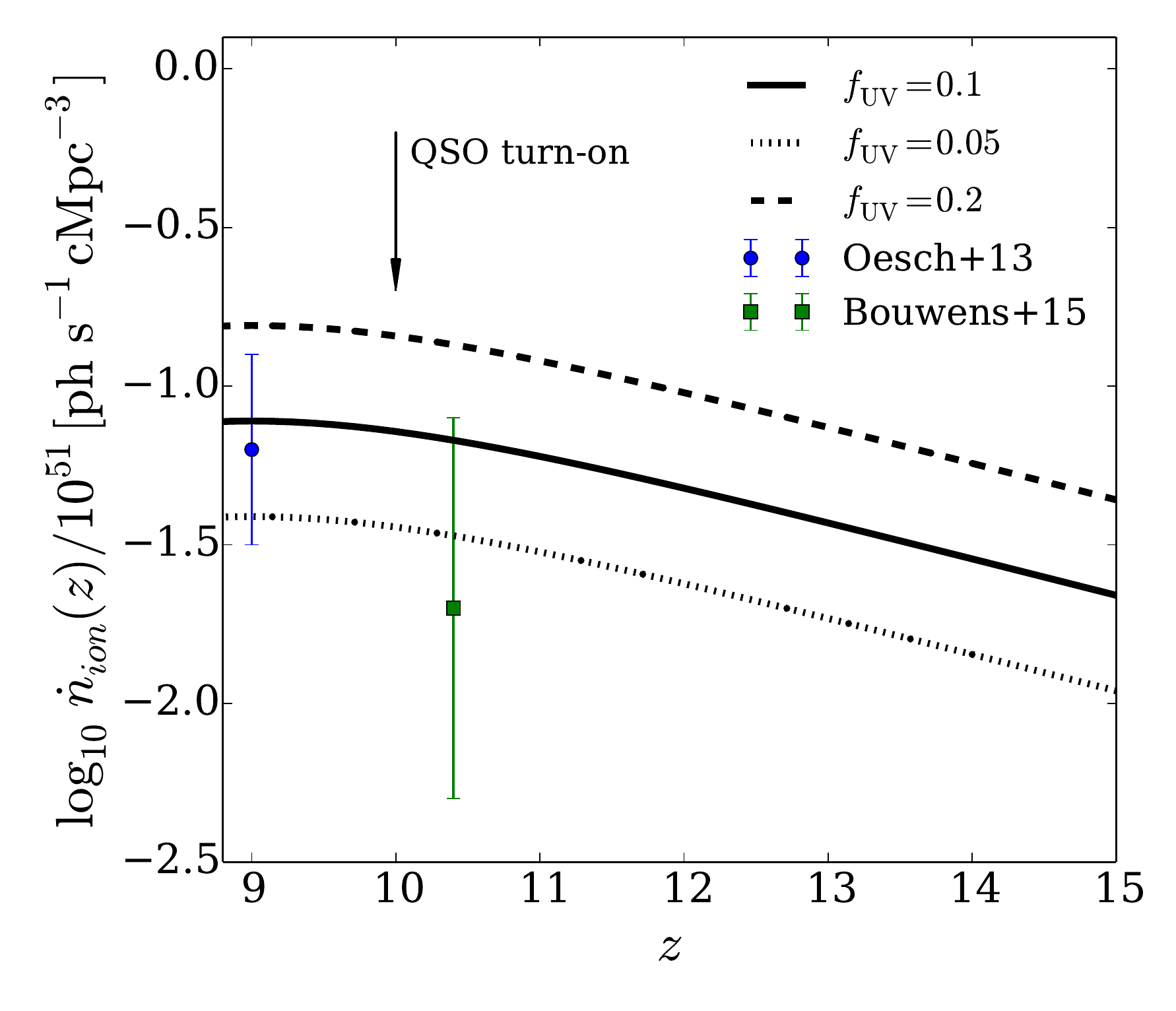}
  \caption[The total ionizing photon emissivity]{Total ionizing comoving emissivity from galaxies as a function of redshift, for $f_{\scriptscriptstyle\rm UV}=0.05,~0.1,~0.2$ (dotted, solid, and dashed lines). The data points are from Becker et al. (2015) based on observations of Lyman-break galaxies by Oesch et al. (2013) and Bouwens et al. (2015). The arrow indicates the redshift ($z=10$) at which a QSO is turned on.}
  \label{fig:ionizing_budget}
\end{figure}

The comoving ionizing photon emissivity is given by (in units of photons~s$^{-1}$ cMpc$^{-3}$),
\begin{equation}
\dot{n}_{ion}(z)=
10^{50.89}\chi(z)\frac{\alpha_b+3}{2\alpha}f_{\scriptscriptstyle\rm UV},
\label{EUV}
\end{equation}
where $\alpha$ and $\alpha_b$ are the extreme-ultraviolet power-law spectral index of the sources and of the ionizing background, respectively, which are assumed to be equal to 3. The redshift dependence of the star formation rate density is parameterized as $\chi(z)=\frac{\xi e^{\zeta(z-9)}}{\xi-\zeta+\zeta e^{\xi(z-9)}}$, with $\xi=14/15$ and $\zeta=2/3$ (\citealt{2007MNRAS.382..325B,2012MNRAS.423..558C}). The emissivity is normalized using the free parameter $f_{\scriptscriptstyle\rm UV}$, which is 0.1 in our fiducial model. Figure~\ref{fig:ionizing_budget} shows a comparison between our ionizing photon emissivity and the values inferred from observations of Lyman-break galaxies assuming an escape fraction of 30 per cent (\citealt{2015PASA...32...45B} based on \citealt{2013ApJ...773...75O} and \citealt{2015ApJ...803...34B}). To take into account the large uncertainties in the value of the emissivity, we have also considered $f_{\scriptscriptstyle\rm UV}=0.05$ and 0.2. 

To distribute the total ionizing photons among galaxies, we assume that each halo hosts one galaxy $i$, whose ionizing photon production rate, $\dot{N}_{ion}^{\rm GAL}(M_{h,i})$ (in units of $\rm photons~s^{-1}$), scales linearly with the host halo mass, i.e.:
\begin{equation}
 \dot{N}_{ion}^{\rm GAL}(M_{h,i})=\dot{n}_{ion}(z)V_{box}\frac{M_{h,i}}{\sum_{j=1}^{N_s}M_{h,j}},
\end{equation}
where $V_{box}$ is the comoving volume of the simulation box and $N_s$ is the total number of galaxies (or haloes) in the simulation. $\dot{n}_{ion}(z)V_{box}$ is the total number of ionizing photons present at redshift $z$.

\begin{figure}
  \centering
  \includegraphics[angle=0,width=\columnwidth]{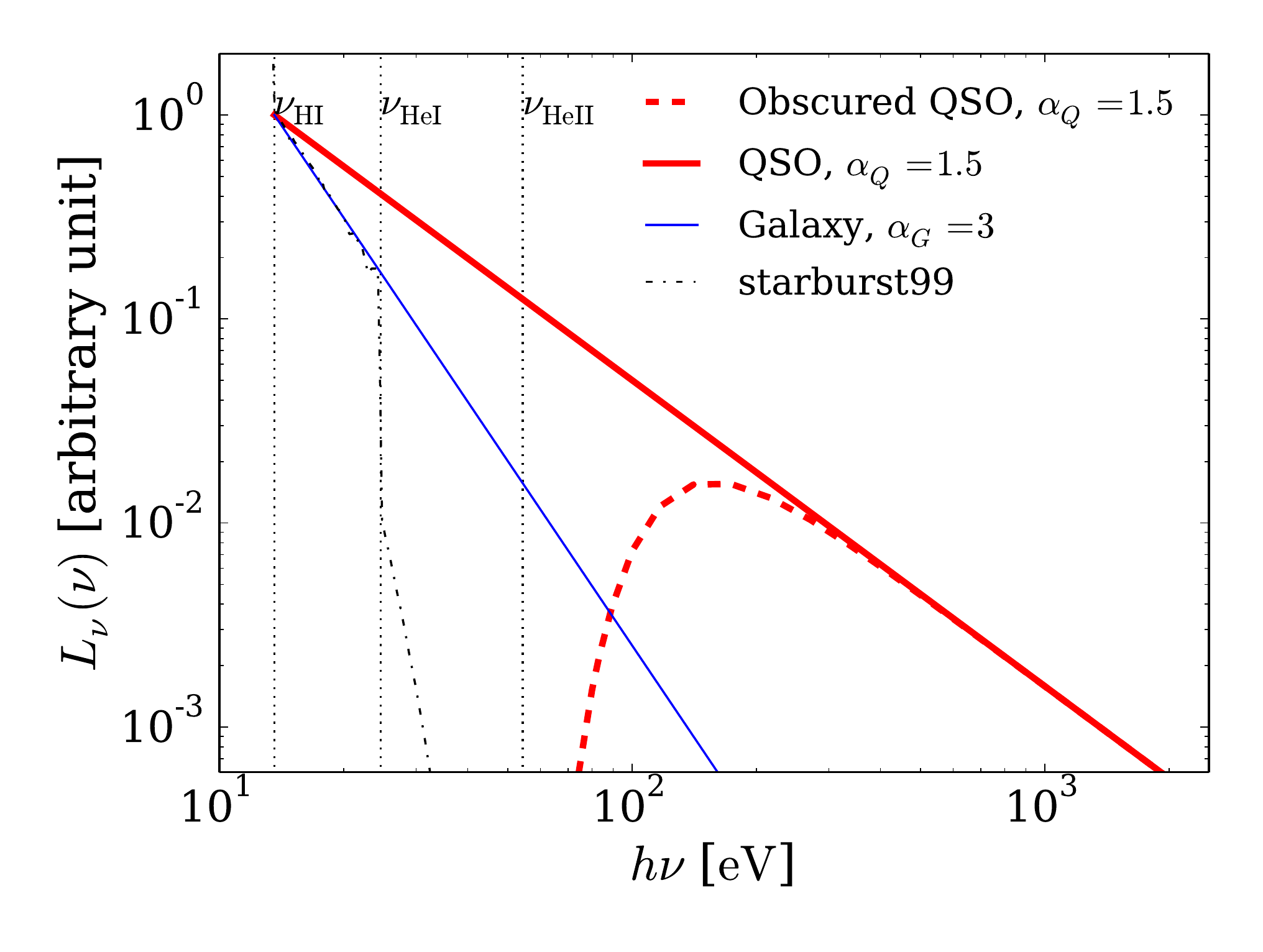}
  \caption[Model spectra of galaxies and a QSO]{Model spectra of a QSO (red thick solid line), obscured QSO (red thick dashed), and galaxy (blue thin solid). For a comparison, the black dash-dotted line shows a spectrum at $10\rm~Myr$ after a burst of star formation with metallicity $Z=0.001$ from starburst99. The vertical lines indicate the ionization thresholds of $\HI$, $\HeI$, and $\HeII$, from left to right. 
}
  \label{fig:spectra}
\end{figure}

The spectrum of a galaxy is modelled with a power-law spectral energy distribution (SED), $L_\nu^{\rm GAL}(\nu)\propto \nu^{-\alpha_G}$, with spectral index $\alpha_G=3$, and it is related to the ionizing photon production rate as $\dot{N}_{ion}^{\rm GAL}=\int_{\nu_L}^\infty (L_\nu^{\rm GAL}(\nu)/h\nu)d\nu$, where $\nu_L=3.29\times10^{15}\rm~Hz$ is the frequency at the Lyman limit. The spectrum is plotted in Figure \ref{fig:spectra} (solid thick blue line), together with a spectrum at $10\rm~Myr$ after a starburst with metallicity $Z=0.001$ from starburst99 (black dotted line; \citealt{1999ApJS..123....3L}). While the SED between $h\nu_{\rm HI}$ and $h\nu_{\rm HeI}$ is well approximated by the power-law, it drops abruptly at $>h\nu_{\rm HeI}$ due to the absorption by the stellar atmosphere in the spectral synthesis model. We have nevertheless adopted a simple power-law for a more straightforward comparison with a QSO model (\S~\ref{qso_source}), and show the $\HeIII$ results for galaxies for completeness. In this paper, we have drawn conclusions that are not affected by this assumption, and we intend to employ a more accurate population synthesized galaxy spectra in future work.

\begin{figure*}
\centering
  \includegraphics[angle=0,width=\textwidth]{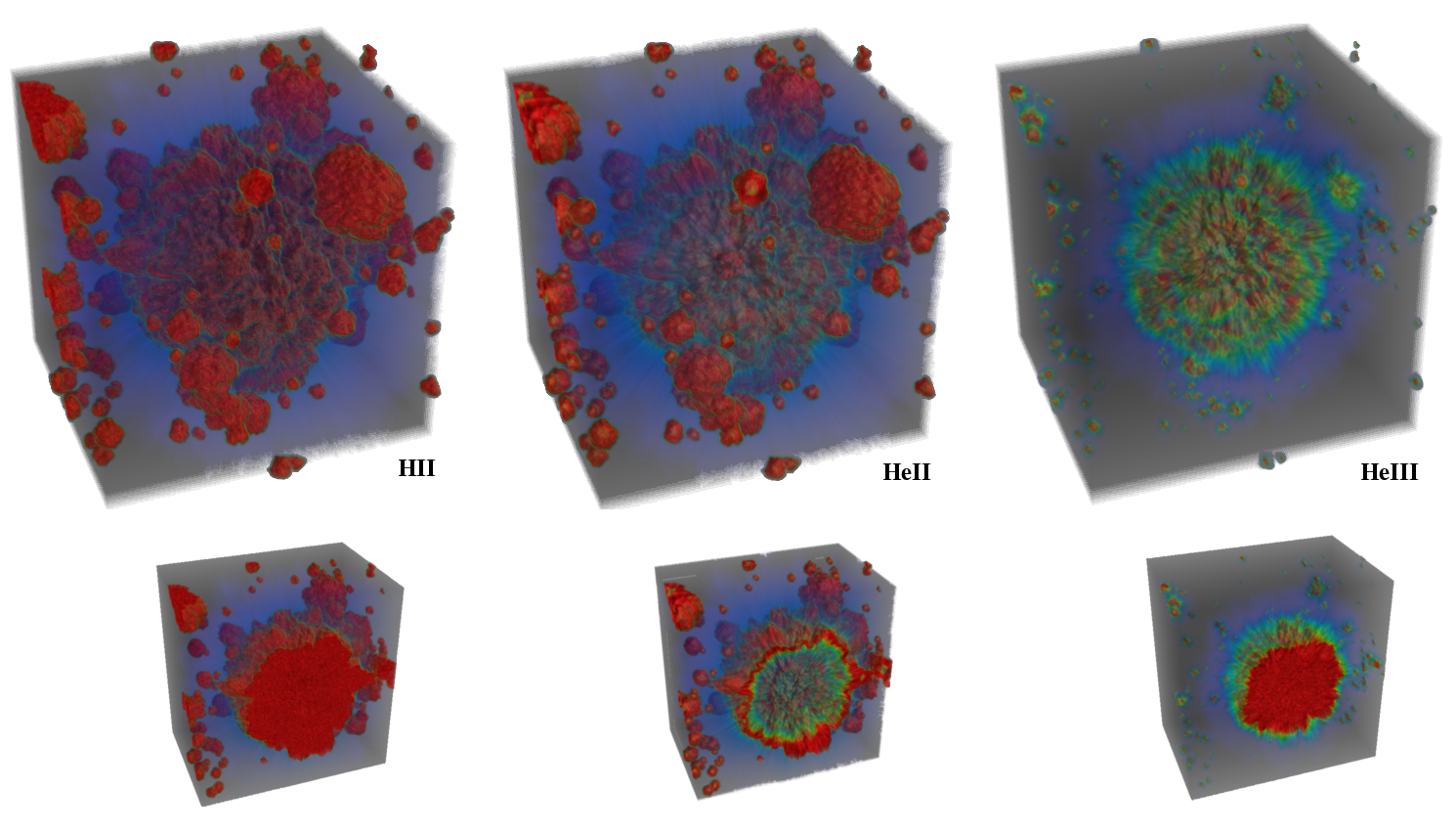}
  \caption[3D visualization of the $\HII$, $\HeII$, and $\HeIII$ fractions]{3D visualization of the $\HII$ (left), $\HeII$ (middle), and $\HeIII$ (right) fractions in the QSO environment, as produced by both the central QSO and the surrounding galaxies with full UV and X-ray physics at (GAL\_R+QSO\_UVXsec). The images refer to a time $t_Q=10^7$~yr after the QSO has been turned on at $z=10$. Unless otherwise stated, all the following figures refer to the same time. {\it Top panels}: the whole simulation box of $50h^{-1}\rm cMpc$. {\it Bottom panels}: dissection through the central QSO. Some ionized regions near the edge appear to reside outside the box because of the visualization reason. The colours correspond to an ionized fraction of $x_i\sim1$ (red), $x_i\sim0.1-0.001$ (green to blue), and $x_i\sim0$ (gray), where $i=\HII,\HeII,\HeIII$. A more quantitative analysis is shown in \S~\ref{neutral_run}-\ref{sec:obscQSO}.}\label{3d}
\end{figure*}

\subsection{Source model: QSO}\label{qso_source}

We model the ionizing photon production rate, $\dot{N}_{ion}^{\rm QSO}$, of our $z=10$ QSO by re-scaling the properties of the ULAS J1120+0641 QSO at $z=7.085$, which is estimated to have a central super-massive black-hole (SMBH) of mass $M_{\rm BH}=2\times10^9\rm~M_\odot$ and $\dot{N}_{ion}^{\rm QSO}=1.36\times10^{57}\rm~photons~s^{-1}$ (\citealt{2011Natur.474..616M,2011MNRAS.416L..70B}).
Assuming a smaller SMBH mass of $M_{\rm BH}=2\times10^8\rm~M_\odot$ to reflect a higher redshift environment and adopting an Eddington limit accretion rate, i.e. the bolometric QSO luminosity scaling linearly with the SMBH mass, this translates into $\dot{N}_{ion}^{\rm QSO}\propto M_{\rm BH}$, giving:
\begin{equation}
\dot{N}_{ion}^{\rm QSO}=1.36\times10^{56}\rm~photons~s^{-1}. \label{QSOluminosity}
\end{equation}

The QSO spectrum is assumed to be a power-law $L_{\nu}^{\rm QSO}(\nu)\propto \nu_L^{-\alpha_Q}$ with index $\alpha_Q=1.5$ (the same as the one of ULAS J1120+0641; \citealt{2011Natur.474..616M,2011MNRAS.416L..70B}), and it is related to the ionizing photon production rate as $\dot{N}_{ion}^{\rm QSO}=\int^\infty_{\nu_L}(L_\nu^{\rm QSO}(\nu)/h\nu)d\nu$. 

In addition, we also consider a model (red dashed line in Figure~\ref{fig:spectra}) in which the QSO spectrum is obscured by a spherical shell of neutral gas with a $\HI$ column density\footnote{This $\HI$ column density is chosen based on lower redshift observations of AGNs (\citealt{2005ApJ...635..864L,2003ApJ...598..886U}) and on the value assumed in a similar theoretical work (\citealt{2008MNRAS.385.1561K})} of $\NHI=10^{20}\rm~cm^{-2}$, i.e. 
$L_{\nu}^{\rm QSO,obsc}(\nu)=L_{\nu}^{\rm QSO}(\nu)e^{-\sigma_{\rm HI}(\nu)\NHI}$, where $\sigma_{\rm HI}(\nu)$ is the $\HI$-photoionization cross section \citep{2011piim.book.....D}.
Note that in an obscured QSO model the ionizing photon production rate is reduced to $\dot{N}_{ion}^{\rm QSO,obsc}=4.33\times10^{54}\rm~photons~s^{-1}$ because a large fraction of ionizing photons with $\nu < \nu_{\rm HeII}$ are absorbed before escaping into the IGM. This effectively hardens the spectrum of the QSO, by removing most of its UV photons.

The lifetime of the QSO is assumed to be $t_Q=10^7\rm~yr$ unless otherwise stated. This is chosen from the range of values $10^6{\rm~yr}<t_Q<10^8{\rm~yr}$ which are observationally allowed at lower redshift (\citealt{2004cbhg.symp..169M}). 
We note that, 
because the recombination timescale is much shorter than the QSO lifetime, increasing lifetime and/or luminosity trivially increases the size of ionized regions as $R_{\rm I}\propto(\dot{N}_{ion}^{\rm QSO}t_Q)^{1/3}$. The temperature is expected to be insensitive to luminosity and lifetime, although the radius to which the gas is photo-heated increases with luminosity and lifetime. We discuss the impact of the QSO duty cycle in Appendix~\ref{app:duty} for interested readers.

In our model, both unobscured and obscured QSOs are assumed to emit ionizing photons isotropically. The opening angle of QSOs at
$z>6$ is not well-constrained, but anisotropic emission will
generally produce an anisotropic (possibly bipolar) region of ionization in the
IGM around the QSOs.

\section{Ionization state of the IGM}\label{sec:H_reion}

In this section we present the results from our suite of radiative transfer simulations with galaxies alone (galaxies only models), a QSO alone (QSO only models), and galaxies and a QSO combined (galaxies+QSO models), and we discuss their impact on hydrogen and helium reionization.

The 3D visualization in Figure~\ref{3d} describes the results for our full reference simulation (GAL\_R+QSO\_UVXsec) at the end of the QSO lifetime\footnote{Unless otherwise stated, all our results are shown at a time corresponding to $t_Q=10^7$~yr after the QSO has been turned on at $z=10$.}. The figure shows clearly the very rich morphology of $\HII$ and $\HeII/\HeIII$ regions in the high-$z$ QSO environment. 
The QSO, together with the surrounding galaxies, form a central large $\HII$ region on scales of tens of comoving mega-parsecs and many smaller $\HII$ regions. The X-ray radiation from the QSO also partially ionizes the IGM beyond the sharp $\HII$ I-front. The $\HeII$ regions show a morphology similar to that of the $\HII$ regions, although the $\HeII$ I-front is slightly broader than the $\HII$ I-front because of the larger mean free path of photons near the $\HeI$ ionization threshold. Differently from $\HII$ and $\HeII$, the $\HeIII$ regions produced by galaxies are extremely small\footnote{Note that a non-negligible contribution of galaxies to the $\HeIII$ regions is a result of our power-law spectrum assumption. For population synthesized galaxy spectra, we expect galaxies to produce negligibly small $\HeIII$ region.}  because of the paucity of $\HeII$-ionizing photons emitted by them. On the other hand, the QSO produces a large almost spherical $\HeIII$ region because of the high $\HeII$-ionizing radiation. The $\HeIII$ I-front is even broader due to the larger mean free path of photons near the $\HeII$ ionization threshold. 

In the following, we explore in detail the impact of varying both the source models (of the galaxies and the QSO) and the physical processes of radiative transfer on the intergalactic hydrogen and helium.

\subsection{$\HII$, $\HeII$ and $\HeIII$ regions from galaxies and a QSO in a neutral IGM}\label{neutral_run}

\begin{figure}
  \centering
  \includegraphics[angle=0,width=\columnwidth]{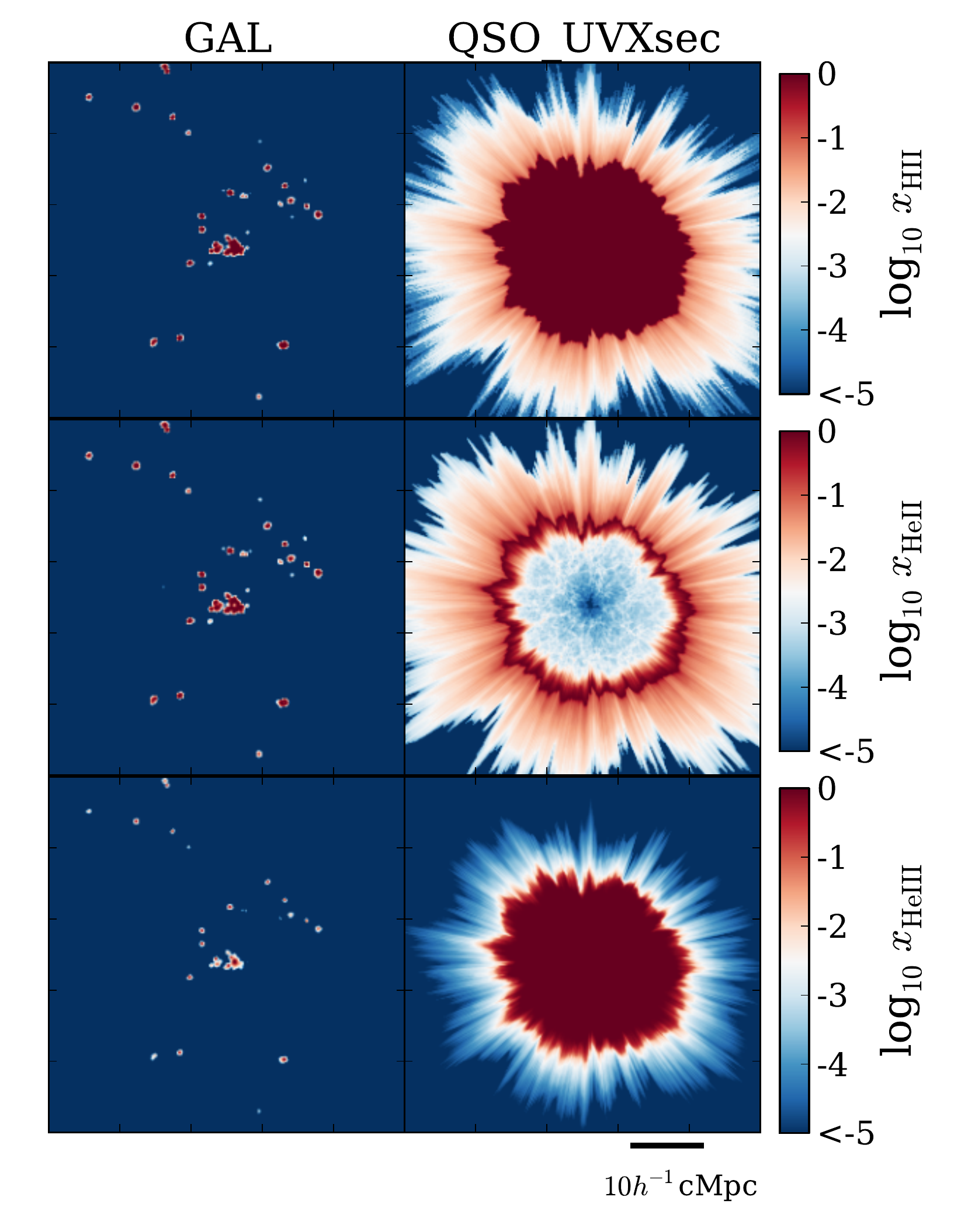}
  \caption[Maps of $\HII$ , $\HeII$ (middle), and $\HeIII$ (bottom) fractions]{Maps of $\HII$ (top panels), $\HeII$ (middle), and $\HeIII$ (bottom) fractions for model GAL (left column) and QSO\_UVXsec (right). In both models the sources are turned on in a fully neutral medium at $z=10$ and shine for $10^7\rm~yr$. The maps have a side length of $50h^{-1}$cMpc and the width of the slice is $195h^{-1}\rm ckpc$.}
  \label{fig:map_neutral}
\end{figure}

We first examine the simplest case in which either galaxies or a QSO create ionized regions in a fully neutral gas. 

Figure \ref{fig:map_neutral} (top panels) shows slices of $\HII$ fraction, $\xHII$, in  models GAL and QSO\_UVXsec, where only galaxies or a QSO are switched on at $z=10$ and for a duration of $10^7\rm~yr$ in a fully neutral IGM. While a single large fully ionized region appears around the QSO, the numerous fainter galaxies produce many smaller $\HII$ regions. This morphological difference is caused by the fact that in the QSO model all the photon budget is concentrated in one single source rather than being distributed among many fainter galaxies. In addition, the QSO emits about $15$ times more ionizing photons than the galaxies. 
Another prominent difference in the ionization structure is the partially ionized shell produced by the X-ray photons emitted by the QSO (see \S~\ref{xray_run} for more details). 

The middle and bottom panels of Figure~\ref{fig:map_neutral} show maps of $\HeII$ and $\HeIII$ fractions. As for the $\HII$ fraction, the GAL and QSO\_UVXsec models present a clearly different morphology.
When reionization is driven by galaxies, the morphology of the $\HeII$ and $\HII$ regions is qualitatively very similar. The reason for this is that, although our galaxy-type spectrum emits a factor of $L_\nu^{\rm GAL}(\nu_{\rm HeI})/L_\nu^{\rm GAL}(\nu_{\rm HI})\simeq0.17$ less $\HeI$- than $\HI$-ionizing photons, the abundance of helium is $Y/X\simeq0.09$ smaller than the one of hydrogen. As a result, when driven by galaxies, the reionization history and morphology of $\HeI$ and $\HI$ are similar. On the other hand, as galaxies emit a factor of $L_\nu^{\rm GAL}(\nu_{\rm HeII})/L_\nu^{\rm GAL}(\nu_{\rm HI})\simeq0.016$ less $\HeII$- than $\HI$-ionizing photons due to the higher ionization threshold of $\HeII$, the $\HeIII$ regions are much more compact.

A QSO, on the other hand, produces a very distinctive $\HeII$ region, as well as a highly regular, large $\HeIII$ region. Because a QSO emits $\simeq 0.4$ $\HeI$-  and $\simeq 0.13$ $\HeII$-ionizing photons per $\HI$-ionizing photon, unlike galasies, a QSO drives $\HeI$ and $\HeII$ reionization simultaneously to $\HI$ reionization, and the $\HeIII$ I-front follows immediately after the $\HeII$ I-front. This creates a `shell'-like structure in $\HeII$, containing a large $\HeIII$ region. Note that this structure  is a distinctive feature in a high-redshift QSO environment; in contrast, at lower redshift ($z<6$) a QSO creates a `hole' in $\HeII$, because $\HeI$ reionization by that time has already been completed by galaxies.

\begin{figure}
\centering
  \includegraphics[angle=0,width=0.85\columnwidth]{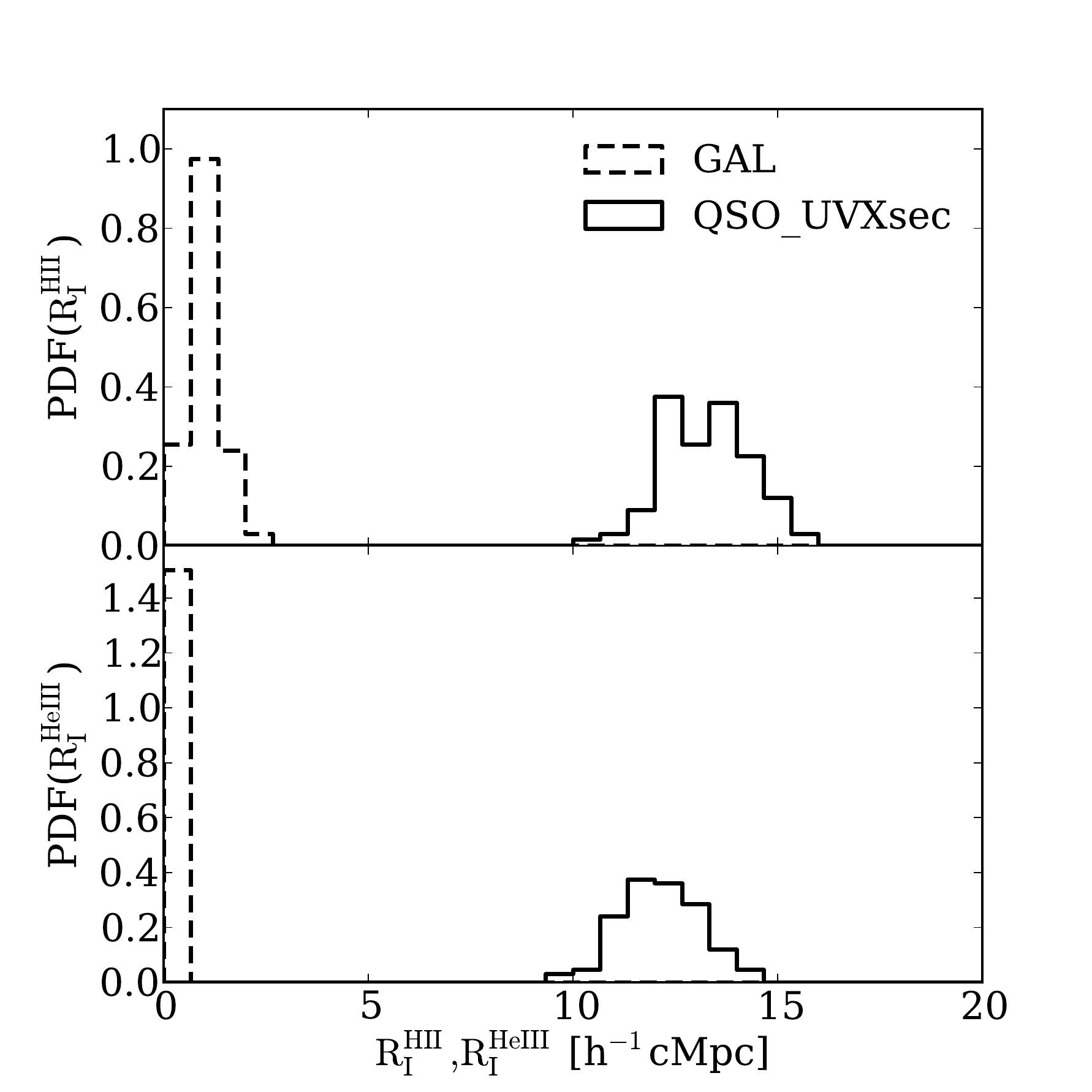}
     \caption[Distribution of the $\HII$ and $\HeIII$ I-front radii]{Distribution of the $\HII$ (top panel) and $\HeIII$ (bottom panel) I-front radii for models GAL (dashed) and QSO\_UVXsec (solid). The PDF of the $\HeII$ I-fronts is nearly identical to that of the $\HII$ I-fronts, and thus it is not shown here.} 
    \label{fig:HIIRadiusPDF1}
\end{figure}

For a more quantitative comparison, in the top panels of Figure~\ref{fig:HIIRadiusPDF1} we plot the distribution of the I-front radii\footnote{The I-front radius is defined as the distance from the central source at which $\xHII$ drops below 50\% for the first time.}, $R^{\rm HII}_{\rm I}$, produced by the QSO (solid line) and the galaxies (dashed line) along 100 random lines-of-sight (LOSs) from the central source. While a central QSO alone can create an $\HII$ region as large as $R^{\rm HII}_{\rm I}\sim 13h^{-1}\rm cMpc$, galaxies produce a smaller I-front with radius $R^{\rm HII}_{\rm I}<3h^{-1}\rm cMpc$.
Furthermore, a QSO-type source shows a dispersion of the $\HII$ I-front radii of $\sim5h^{-1}\rm cMpc$. This is caused by density fluctuations of the IGM: while dense gas clumps cast shadows behind them, the I-fronts propagate unimpeded along underdense directions (\citealt{2006MNRAS.371.1057I,2015MNRAS.454..681K} and also see Graziani et al., in prep.).  

The distribution of the $\HeIII$ I-front radii is shown in the bottom panel of Figure~\ref{fig:HIIRadiusPDF1}. QSO's $\HeIII$ I-fronts (solid line) have an extension of $R^{\rm HeIII}_{\rm I}\sim 12h^{-1}\rm cMpc$, slightly smaller than the corresponding $\HII$ I-fronts. Galaxies (dashed line) produce negligibly small radii. The dispersions of the QSO's $\HII$ and $\HeIII$ I-fronts are very similar, as they are both caused by shadowing by dense clumps. 

\subsection{Effect of X-rays and secondary ionization}\label{xray_run}

\begin{figure}
\centering
 \includegraphics[angle=0,width=\columnwidth]{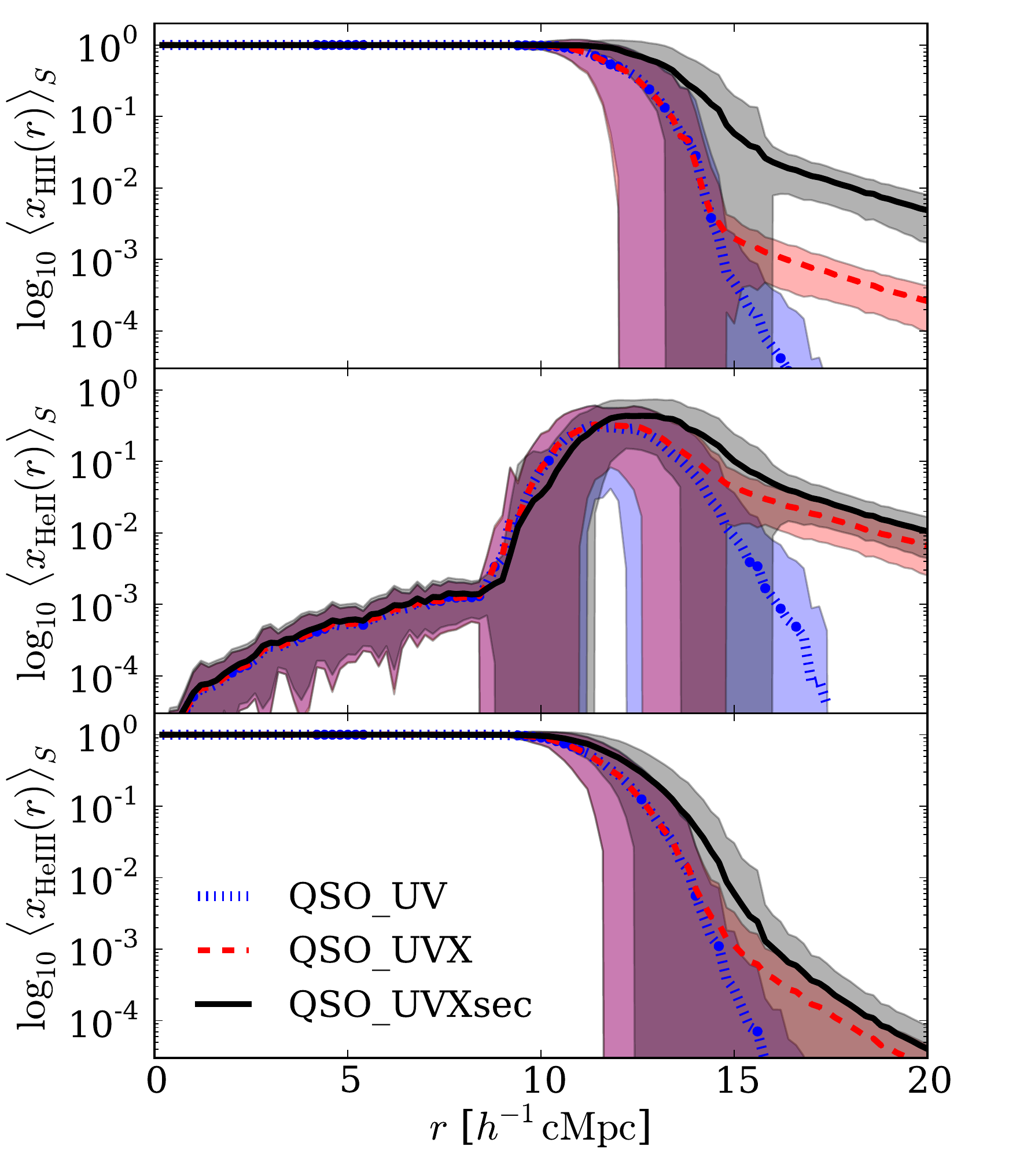}
  \caption[Spherically averaged profiles of $\HII$,  $\HeII$, and  $\HeIII$ fractions]{Spherically averaged profiles of $\HII$ (top panel),  $\HeII$ (middle), and  $\HeIII$ (bottom) fractions around a QSO turned on at $z=10$ in a neutral medium including only UV photons (QSO\_UV, blue dotted lines), UV and X-ray photons (QSO\_UVX, red dashed), UV, X-ray photon and secondary ionization (QSO\_UVXsec, black solid). The shaded regions show the 1$\sigma$ scatter of line-of-sight fluctuations. The profiles are shown at the end of the QSO lifetime $t_Q=10^7$~yr.}
\label{fig:x-ray_radial}
\end{figure}

Here we study the effect of X-rays and secondary ionization on the region ionized by the central QSO. This clarifies the physical origin of the extended tail of partial ionization in the QSO model seen in Figures~\ref{3d} and \ref{fig:map_neutral}.

Figure~\ref{fig:x-ray_radial} shows the spherically averaged profiles of ionization fractions, $\langle x_i(r)\rangle_S$, together with the corresponding $1\sigma$ scatter, $\sigma_i^2(r)=\langle x_i^2(r)\rangle_S-\langle x_i(r)\rangle_S^2$, where $i=\HII,\HeII,\HeIII$. The figure compares the three QSO only models with different RT processes, i.e. spectra restricted only to UV photons (QSO\_UV), including UV+X-ray photons but neglecting the effect of secondary ionization (QSO\_UVX), and including UV+X-ray photons and secondary ionization (QSO\_UVXsec). The main impact of X-rays and secondary ionization on both hydrogen and helium is to produce an extended tail of low ionization beyond the I-fronts. 

In the case of hydrogen, while the impact of X-rays is to increase the ionized fraction to $\langle \xHII(r)\rangle_S\sim10^{-3}$ (QSO\_UVX), secondary ionization further enhances it to $\langle \xHII(r)\rangle_S\sim10^{-2}$ (QSO\_UVXsec) (see also \citealt{2011MNRAS.414.3458W}). The ionized fraction has line-of-sight fluctuations of about a factor of $\sim3$ as a result of density fluctuations\footnote{The increase in scatter observed near the I-front is due to large variations in the I-front position (between $\sim 10h^{-1}\rm cMpc$ and $\sim 15h^{-1}\rm cMpc$), rather than to the effect of X-rays.}.  

Similarly, the partially ionized tails of $\HeII$ and $\HeIII$ have ionized fractions of $\langle\xHeII(r)\rangle_S\sim10^{-2}$ and $\langle\xHeIII(r)\rangle_S\sim10^{-4}-10^{-3}$. However, the impact of secondary ionization is less significant than on hydrogen, with an enhancement of only a factor of $\sim2$. This is because, due to its lower abundance, the probability that helium is collisionally ionized by fast electrons is, to a first order approximation, a factor of $Y/X\simeq0.09$ lower than the one for hydrogen. 

\begin{figure}
\centering
 \includegraphics[angle=0,width=0.52\columnwidth]{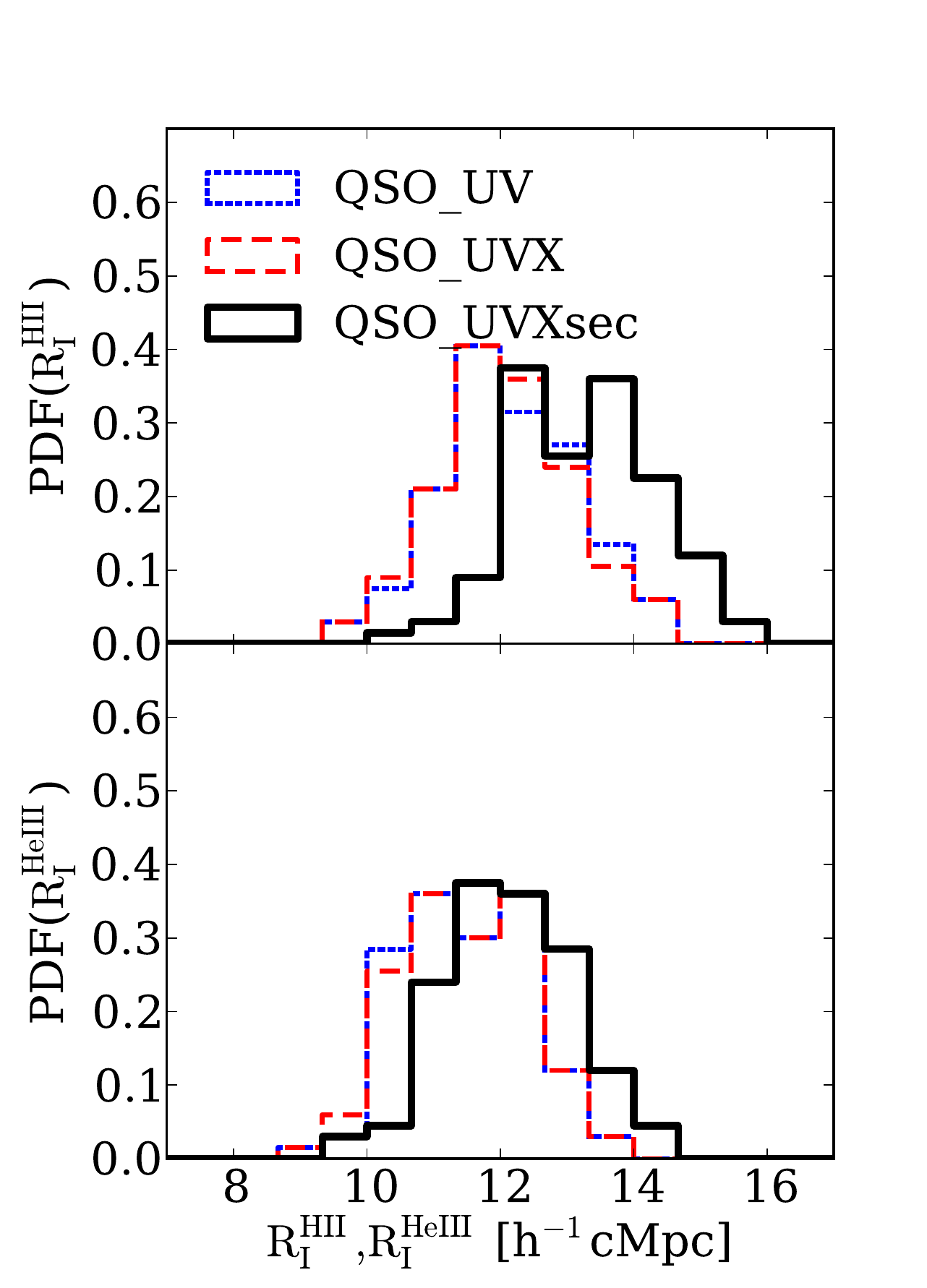}
\hspace{-0.5cm}
 \includegraphics[angle=0,width=0.52\columnwidth]{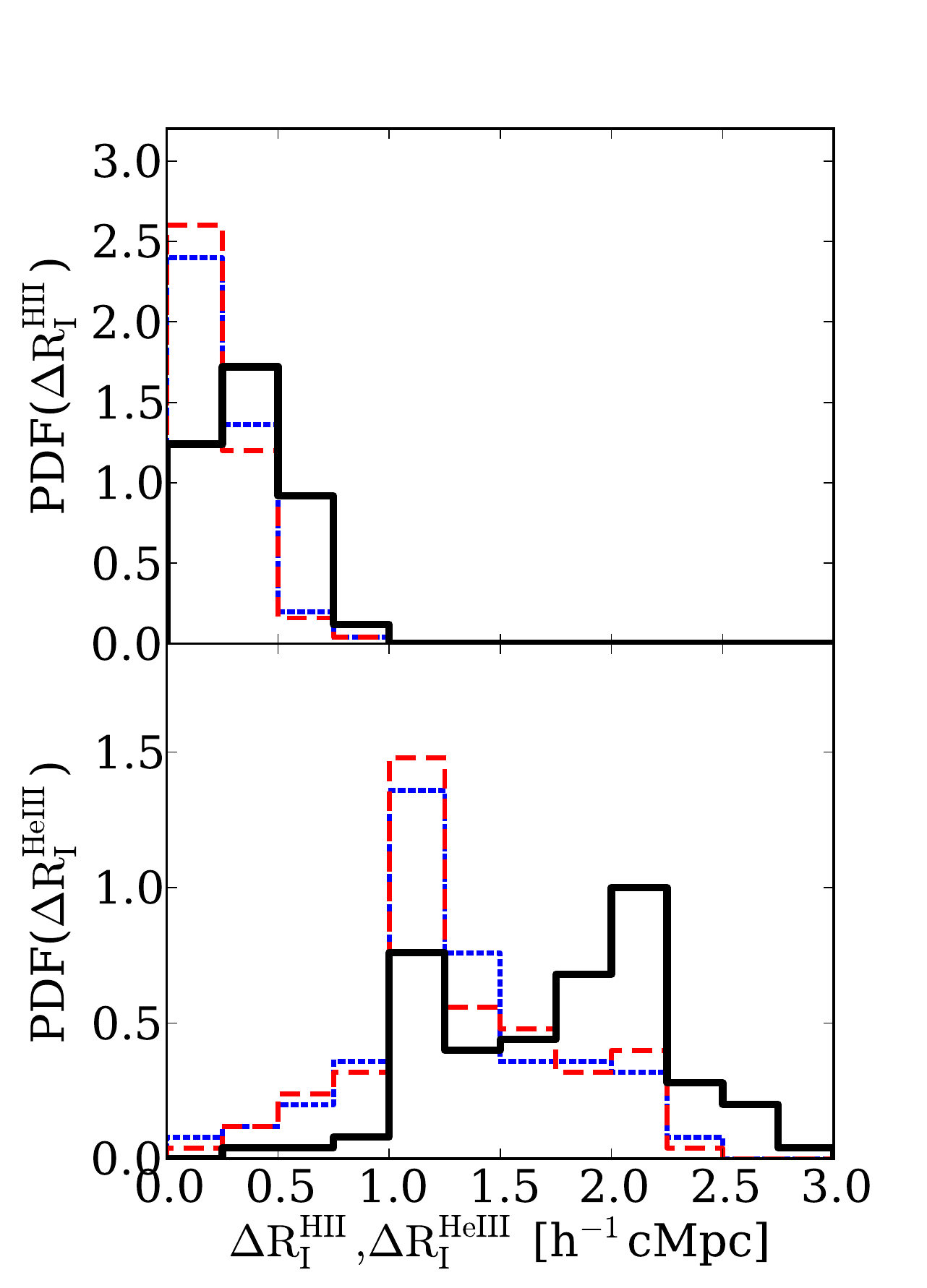}
  \caption[Distribution of the $\HII/\HeIII$ I-front position and thickness]{Distribution of the $\HII/\HeIII$ I-front positions (left panels) and thickness (right panels) for the three QSO models including only UV photons (QSO\_UV, blue dotted), UV and X-ray photons (QSO\_UVX, red dashed), UV, X-ray photon and secondary ionization (QSO\_UVXsec, black solid).}
\label{fig:x-ray_hist}
\end{figure}

\begin{figure*}
  \includegraphics[angle=0,width=1.05\textwidth]{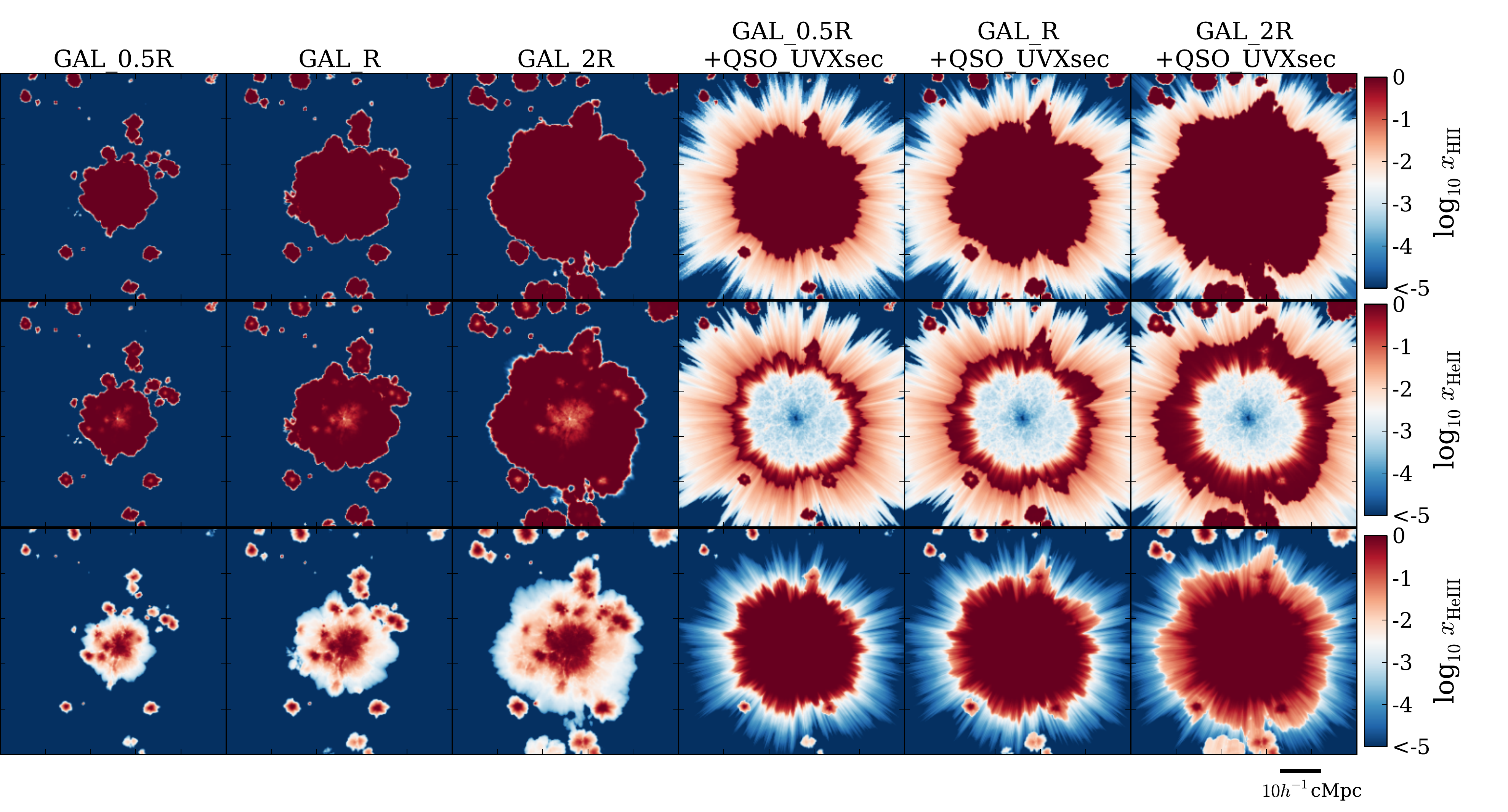}
  \caption[Maps of $\HII/\HeII/\HeIII$ fractions in galaxies only and galaxies+QSO models]{Maps of $\HII$ (top panels), $\HeII$ (middle), and $\HeIII$ (bottom) fraction for galaxies only models GAL\_0.5R, GAL\_R, GAL\_2R, and galaxies+QSO models GAL\_0.5R+QSO\_UVXsec, GAL\_R+QSO\_UVXsec, GAL\_2R+QSO\_UVXsec. In both sets of models, galaxies start to reionize the IGM at $z=15$, while the QSO is turned on at $z=10$ for a duration of $10^7\rm~yr$ in the medium already reionized by the pre-existing galaxies. The maps have a side length of $50h^{-1}\rm cMpc$ and the width of a slice is $195h^{-1}\rm ckpc$.}
  \label{fig:reion_HII}
\end{figure*}

Figure~\ref{fig:x-ray_hist} shows the distribution of the $\HII$ and $\HeIII$ I-front radii and their I-front thickness\footnote{The thickness is defined as the distance between the location where the ionized fraction is $90\%$ and that at which it becomes $10\%$. As a first approximation, the thickness is about a mean free path, $\lambda_{\rm mfp}(\nu)=[\nH(z)\sigma_{\HI}(\nu)]^{-1}$, of typical photons reaching the I-front (e.g. \citealt{1998ppim.book.....S,2005MNRAS.360L..64Z,2008MNRAS.385.1561K}). Although the thickness of the I-front is unresolved, i.e. $\lambda_{\rm mfp}<\Delta x_{\rm cell}$ where $\Delta x_{\rm cell}=195h^{-1}\rm kpc$ is the cell size of our simulations, we use the above definition to support the claim that the presence of X-ray photons alone does not modify the thickness of the I-fronts.} based on 100 random LOSs for the three QSO models discussed above. When neglecting secondary ionization, both distributions are weakly affected by X-rays (red dashed vs blue dotted lines). This is a direct consequence of the fact that most of the $\HI$ and $\HeII$-ionizing power is dominated by photons emitted near the ionization thresholds, i.e. the UV photons.  
The photo-ionization by X-ray photons alone does not sensitively modify the ionization state of the IGM near the I-fronts, while the extended tail of partially ionized regions is indeed a strong feature of X-rays.

With secondary ionization, the I-front positions are pushed slightly outward, by $\sim1h^{-1}\rm cMpc$, and their thickness is broadened by $\sim50$ per cent. While this could be a result of secondary ionization preferentially enhancing the ionization level in initially more neutral regions, we caution that the extent of such effect is also related to the gas distribution and the model used for the physics of secondary ionization (Graziani et al., in prep.), as well as the I-front definition.

\subsection{$\HII$, $\HeII$ and $\HeIII$ regions from galaxies and a QSO in a pre-ionized IGM}\label{reion_run}

We now analyse the more realistic case in which a QSO is turned on at $z=10$ in a medium which has already been ionized by pre-existing galaxies.  
We consider three different reionization histories, which differ only for the total ionizing photon emissivity, namely models 2R, R, and 0.5R. As a reference, the $\HII$ ($\HeII$) volume-averaged fraction at $z=10$ is $\langle\xHII\rangle_V\approx28\%,~13\%,~6\%$ ($\langle\xHeII\rangle_V\approx26\%,~12\%,~5\%$) for 2R, R, 0.5R, respectively. 
On the other hand, the contribution of the pre-existing galaxies to $\HeII$ reionization is negligibly small.
In the following, we analyse in great detail how the combined effect of a QSO and galaxies shape the ionization state of the IGM.

The top panels of Figure~\ref{fig:reion_HII} show maps of $\HII$ fraction for both galaxies only and galaxies+QSO models. 
Differently from the scenario in \S~\ref{neutral_run}, the combined effect of galaxies and a QSO produces a morphology (size and shape) of the fully ionized $\HII$ regions (with $\xHII\gtrsim0.9$) similar to the one when no QSO is present (see Figure 9 for a quantitative description). Galaxies alone, in fact, can also produce a large quasi-spherical $\HII$ region due to a combination of their clustering in the high density regions where QSOs preferentially reside, and the long timescale which they have been shining for, allowing for the growth and overlap of many small $\HII$ regions. 
The QSO contributes to produce a slightly larger $\HII$ region (see discussion below) and a prominent tail of partial ionization, which remains a distinctive feature produced by the X-ray photons (see \S~\ref{xray_run}).

 \begin{figure}
 \centering
   \includegraphics[angle=0,width=0.515\columnwidth]{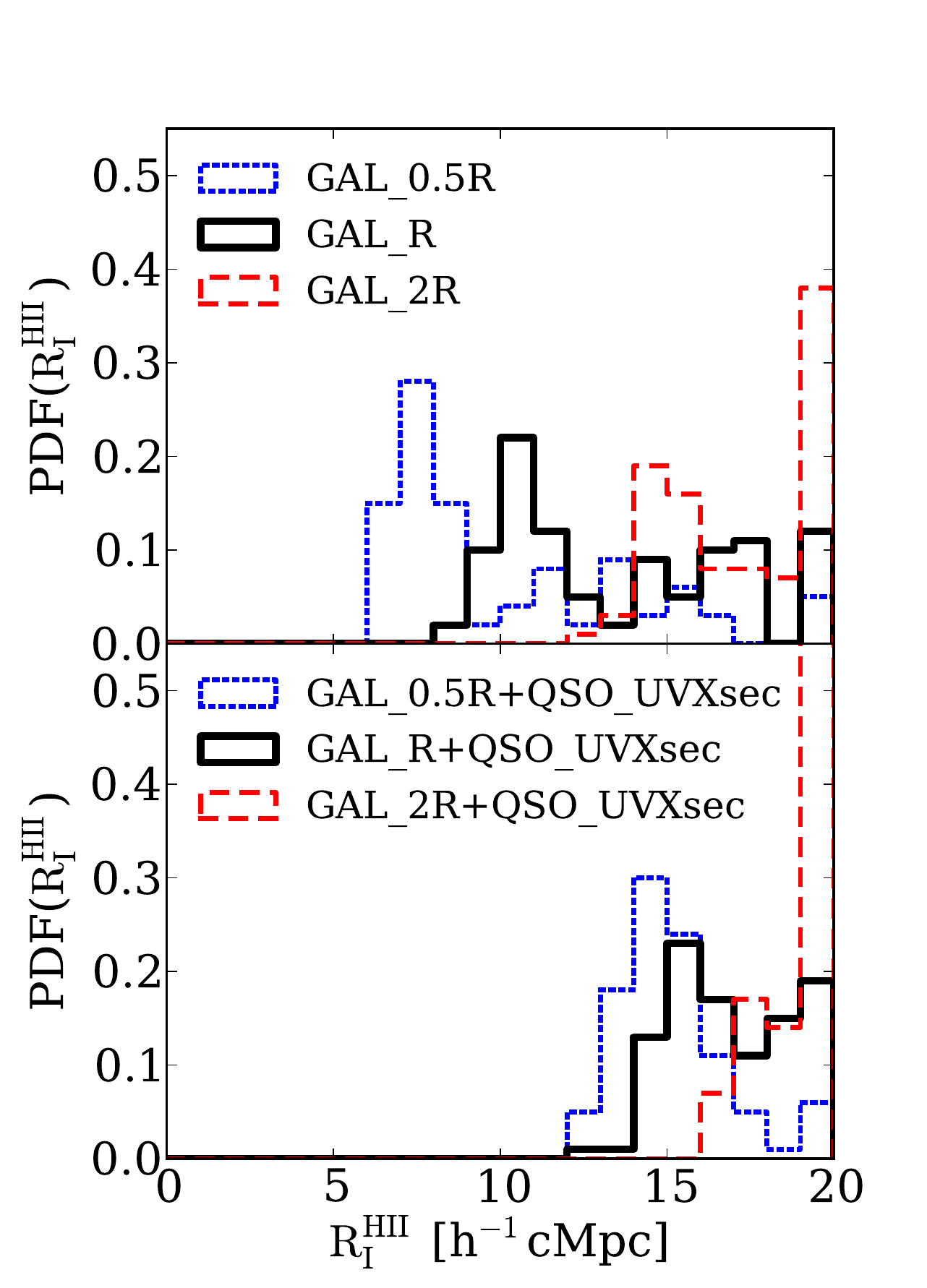}
 \hspace{-0.44cm}
   \includegraphics[angle=0,width=0.515\columnwidth]{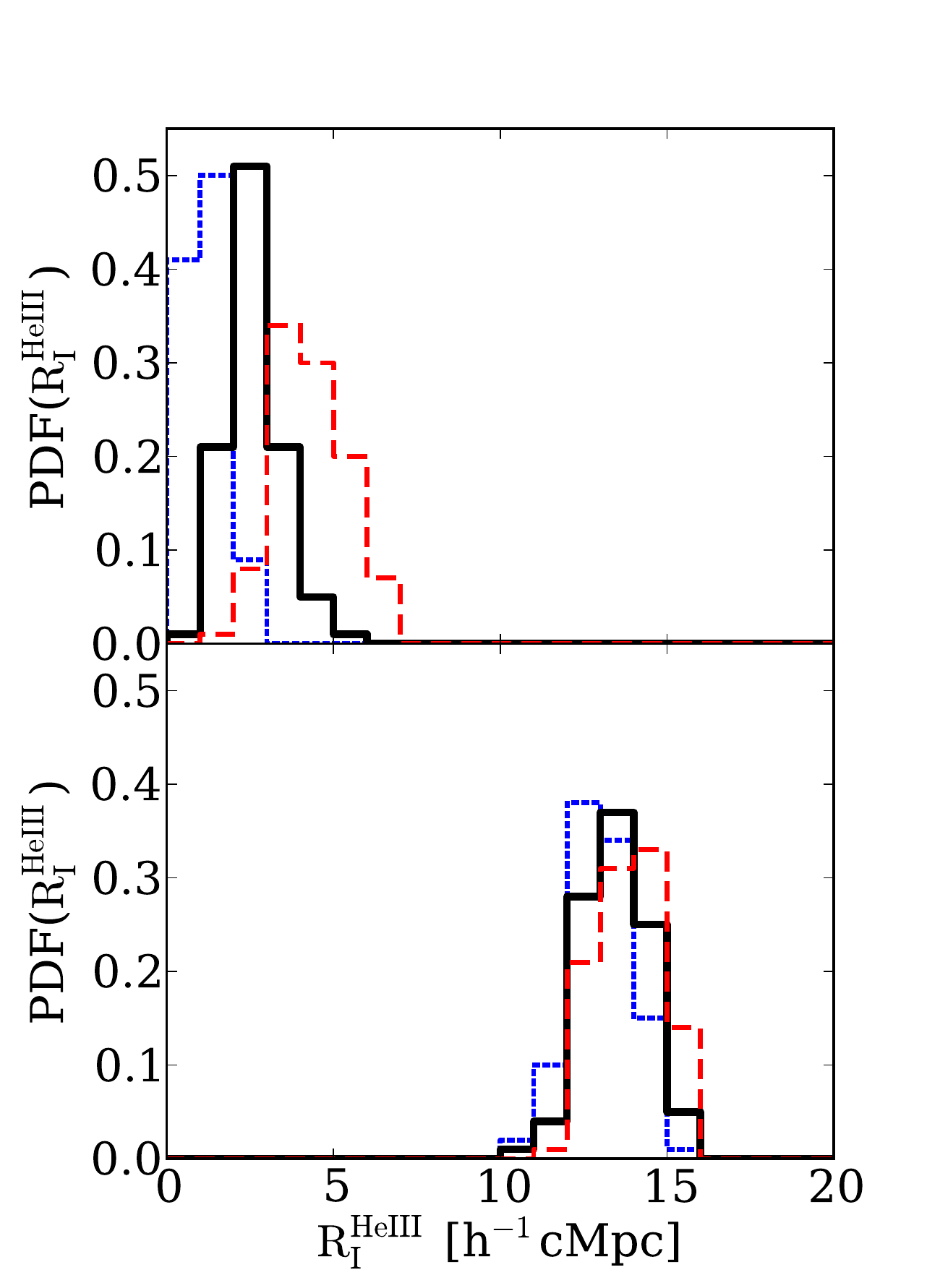}
   \caption{Distribution of the $\HII$ (left panels) and $\HeIII$ (right panels) I-front positions for galaxies only (top) and galaxies+QSO (bottom) models. The linestyle indicates the different source models of galaxies (i.e. different total ionizing photon emissivity): GAL\_0.5R (blue dotted), GAL\_R (black solid), and GAL\_2R (red dashed).}
   \label{fig:HIIRadiusPDF2}
 \end{figure}

The middle and bottom panels of Figure~\ref{fig:reion_HII} show the corresponding maps of $\HeII$ and $\HeIII$ fractions. Unlike for the $\HII$ regions, the galaxies only and galaxies+QSO models exhibit very distinctive features not only in the partially ionized tails, but also in the fully ionized $\HeII/\HeIII$ regions.
In the galaxies only models $\HeI$ and $\HI$ reionization progress simultaneously and have a similar morphology. The QSO just increases the extent of the pre-existing $\HeII$ regions. At the same time, its hard spectrum doubly ionizes helium behind the $\HeII$ I-fronts induced by galaxies, producing an extended $\HeIII$ region, and leaving only a shell of $\HeII$. The width of the shell increases with the ionizing photon emissivity of galaxies because the extent of the pre-existing $\HeII$ region is larger. As the QSO dominates the ionization of $\HeII$ to $\HeIII$, the $\HeIII$ region is very regular, almost spherical.

The left panels of Figure~\ref{fig:HIIRadiusPDF2} show the distribution of the $\HII$ I-front radii for galaxies only and galaxies+QSO models. As expected, the radius increases with the total ionizing photon emissivity. With our I-front definition, the dispersion is $>5h^{-1}\rm~cMpc$, larger than the one in the QSO only models (see Figure \ref{fig:HIIRadiusPDF1}), because it is induced not only by the density fluctuations as discussed in \S~\ref{neutral_run}, but also by the irregularities which follow the overlap of multiple smaller $\HII$ regions.
The QSO pushes the $\HeIII$ I-front to $\sim13h^{-1}\rm cMpc$ (right panels). As the QSO dominates the production of the $\HeIII$ region, the I-front radius is insensitive to the change in total ionizing emissivity of galaxies. In addition, the dispersion of the $\HeIII$ I-front is limited to only $\sim5h^{-1}\rm cMpc$, smaller than that of the $\HII$ I-front, because the fluctuations in this case are caused only by the density fluctuations of the IGM (as discussed in \S~\ref{neutral_run}), while the irregularities due to the overlap of negligibly small $\HeIII$ regions produced by galaxies are irrelevant.

\subsubsection{Relative importance of galaxies and QSO}\label{sec:clustering}

We highlight that the ratio of total cumulative ionizing photons emitted by galaxies and QSOs since the onset of reionization is the most important quantity for determining the ionization structure of the IGM during reionization. We elaborate on the role of galaxies around a QSO by the following simple illustrative argument.

The total number of ionizing photons emitted by galaxies within a comoving radius $r$ around a QSO since the onset of reionization, $N^{\rm GAL}_{ion}(<r)$, can be estimated by
\begin{align}
&N^{\rm GAL}_{ion}(<r)=\int_{z}^{z_0}dz'\left|\frac{dt}{dz'}\right|\dot{n}_{ion}(z')\int_0^r dr' 4\pi r'^2[1+\xi_{qg}(r')], \nonumber \\
&~~~~~~~~~~~~~~=\frac{4\pi }{3}n_{ion}^{\rm GAL}(z)r^3\left[1+\frac{4\pi}{3-\gamma}\left(\frac{r}{r_0}\right)^{-\gamma}\right],\label{eq6}
\end{align}
where $n_{ion}^{\rm GAL}(z)=\int_{z}^{z_0}dz'\left|dt/dz'\right|\dot{n}_{ion}(z')$, with $\left|dt/dz\right|=1/[H(z)(1+z)]$, is the total comoving ionizing photon density from galaxies between the onset of reionization $z_0$ and redshift $z$. $\xi_{qg}(r)=(r/r_0)^{-\gamma}$ is the QSO-galaxy correlation function, where $r_0$ is the correlation length and $\gamma$ is the power-law index. The I-front radius of the $\HII$ region produced by such population of galaxies, $R_{\rm I,gal}^{\rm HII}$, can be estimated by equating the number of ionizing photons and hydrogen atoms inside the radius, i.e. $(4\pi/3)\bar{n}_{\rm H}(0)(R_{\rm I,gal}^{\rm HII})^3=N_{ion}^{\rm GAL}(<R_{\rm I,gal}^{\rm HII})$, obtaining
\begin{equation}
R_{\rm I, gal}^{\rm HII}=r_0\left[\frac{3-\gamma}{4\pi}\left(\frac{\bar{n}_{\rm H}(0)}{n_{ion}^{\rm GAL}(z)}-1\right)\right]^{-1/\gamma},
\end{equation}
where $\bar{n}_{\rm H}(0)$ is the mean comoving hydrogen number density. Using the best-fit power-law QSO-galaxy correlation function to our simulation, i.e. $r_0=18.6h^{-1}\rm cMpc$ and $\gamma=1.6$, we find $R_{\rm I,gal}^{\rm HII}\approx12h^{-1}\rm cMpc$ for our reference galaxies model (GAL\_R), which is in reasonable agreement with the simulation (see Figure~\ref{fig:HIIRadiusPDF2}).

Once a QSO is switched on in the $\HII$ region produced by pre-existing galaxies, it dominates the growth of the I-front radius, $R_{\rm I}^{\rm HII}$, which can be estimated from
(e.g. \citealt{1987ApJ...321L.107S,1999ApJ...514..648M})
\begin{equation}
 \frac{4\pi}{3}(R_{\rm I}^{\rm HII})^3-\frac{4\pi}{3}(R_{\rm I, gal}^{\rm HII})^3
 =\frac{\dot{N}_{ion}^{\rm QSO}\bar{t}_{\rm rec, HII}}{\bar{n}_{\rm H}(0)}\left(1-e^{-t_Q/\bar{t}_{\rm rec, HII}}\right),
\end{equation}
where $\bar{t}_{\rm rec, HII}$ is the recombination timescale of hydrogen.
Assuming that the QSO lifetime is much shorter than the recombination timescale, the final I-front radius is given by 
\begin{equation}
R_{\rm I}^{\rm HII}=R_{\rm I, gal}^{\rm HII}\left[1+ \left(\frac{R_{\rm I, QSO}^{\rm HII}}{R_{\rm I, gal}^{\rm HII}}\right)^3\right]^{1/3},
\label{preHIIStromgren}
\end{equation}
where $R_{\rm I, QSO}^{\rm HII}=\left[\frac{3\dot{N}_{ion}^{\rm QSO}t_Q}{4\pi \bar{n}_{\rm H}(0)}\right]^{1/3}$ is the comoving I-front radius produced by a QSO when it is turned on in isolation. For our reference values we find $R_{\rm I, QSO}^{\rm HI}\approx11.8h^{-1}\rm cMpc$, consistent with the results of the simulations shown in Figure \ref{fig:HIIRadiusPDF1} (QSO\_UVXsec). Finally, the total I-front radius including both galaxies and a central QSO is $R_{\rm I}^{\rm HI}\approx15h^{-1}\rm cMpc$, again in reasonable agreement with the simulations (see Figure~\ref{fig:HIIRadiusPDF2}, GAL\_R+QSO\_UVXsec).

The fact that the simulated size of the $\HII$ region can be broadly obtained using simple analytic estimates indicates that the ionization structure of the IGM largely depends on (\rmnum{1}) the total cumulative ionizing photon emissivity from galaxies, $n_{ion}^{\rm GAL}(z)$, (\rmnum{2}) the QSO-galaxy correlation function, $\xi_{qg}(r)$, and (\rmnum{3}) the ionizing photons emitted by a QSO, $\dot{N}_{ion}^{\rm QSO}t_Q$. More specifically, the final size of the $\HII$ region strongly depends on {\it the ratio between the product of the ionizing photon production rate and the lifetime of the central QSO, $\dot{N}_{ion}^{\rm QSO}t_Q$, and the integrated total ionizing photons of galaxies around a QSO, $N^{\rm GAL}_{ion}(<r)$, since the onset of reionization}. If the cumulative number of photons emitted from galaxies since the onset of reionization exceeds or is comparable to that from a QSO during its activity, the resulting $\HII$ region is very similar. On the other hand, if a QSO dominates the total ionizing photons in the environment, the QSO imprints a distinctive $\HII$ region.

\begin{figure}
  \centering
  \includegraphics[angle=0,width=\columnwidth]{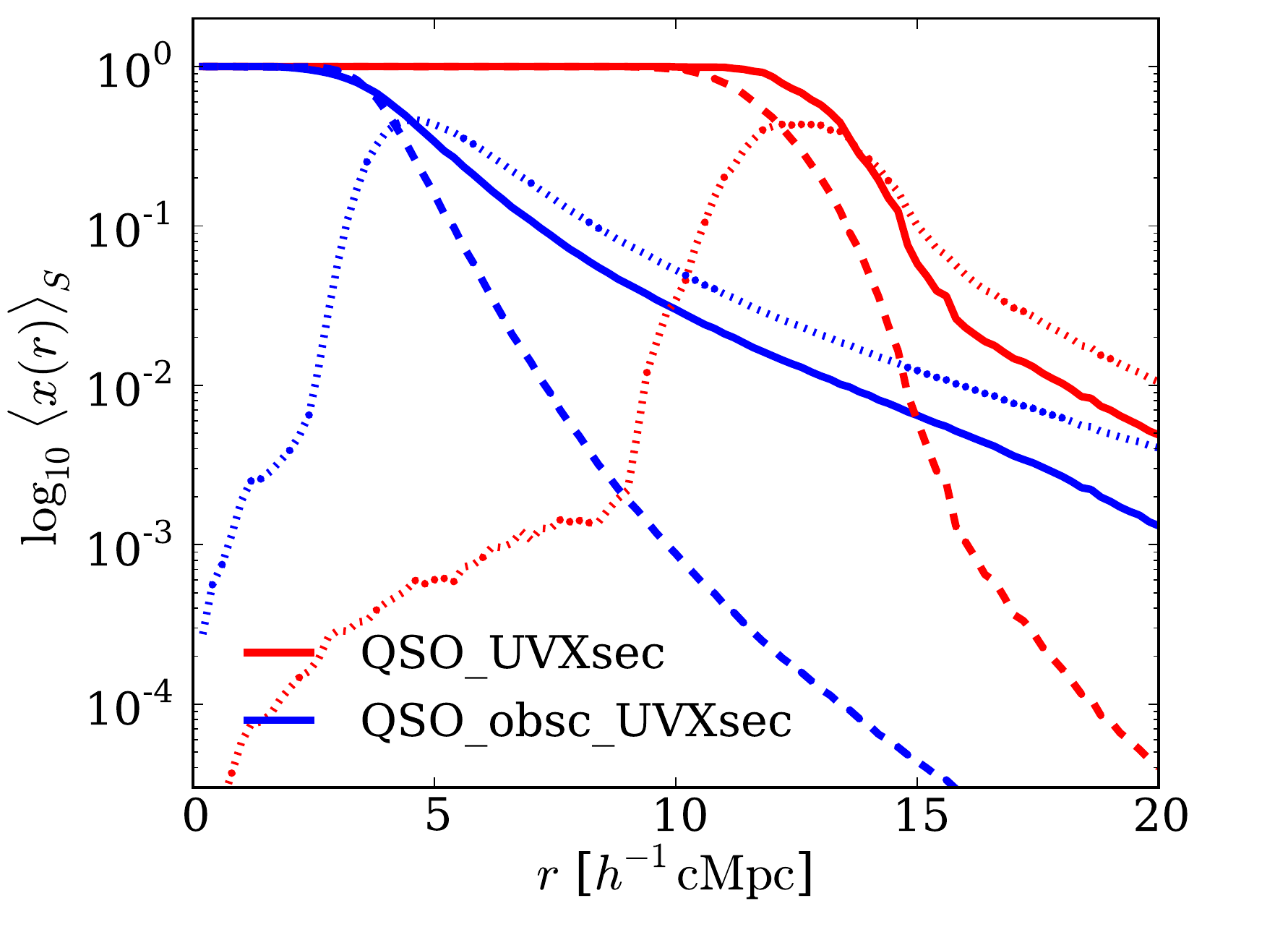}
\caption{Spherically averaged profiles of $\HII$ (solid), $\HeII$ (dotted), and $\HeIII$ (dashed) fractions for QSO (QSO\_UVXsec, red lines) and obscured QSO (QSO\_obsc\_UVXsec, blue lines) models.  } \label{fig:radial_profile_obscured}
\end{figure}

\begin{figure*}
\advance\leftskip+1.cm
  \includegraphics[angle=0,width=\columnwidth]{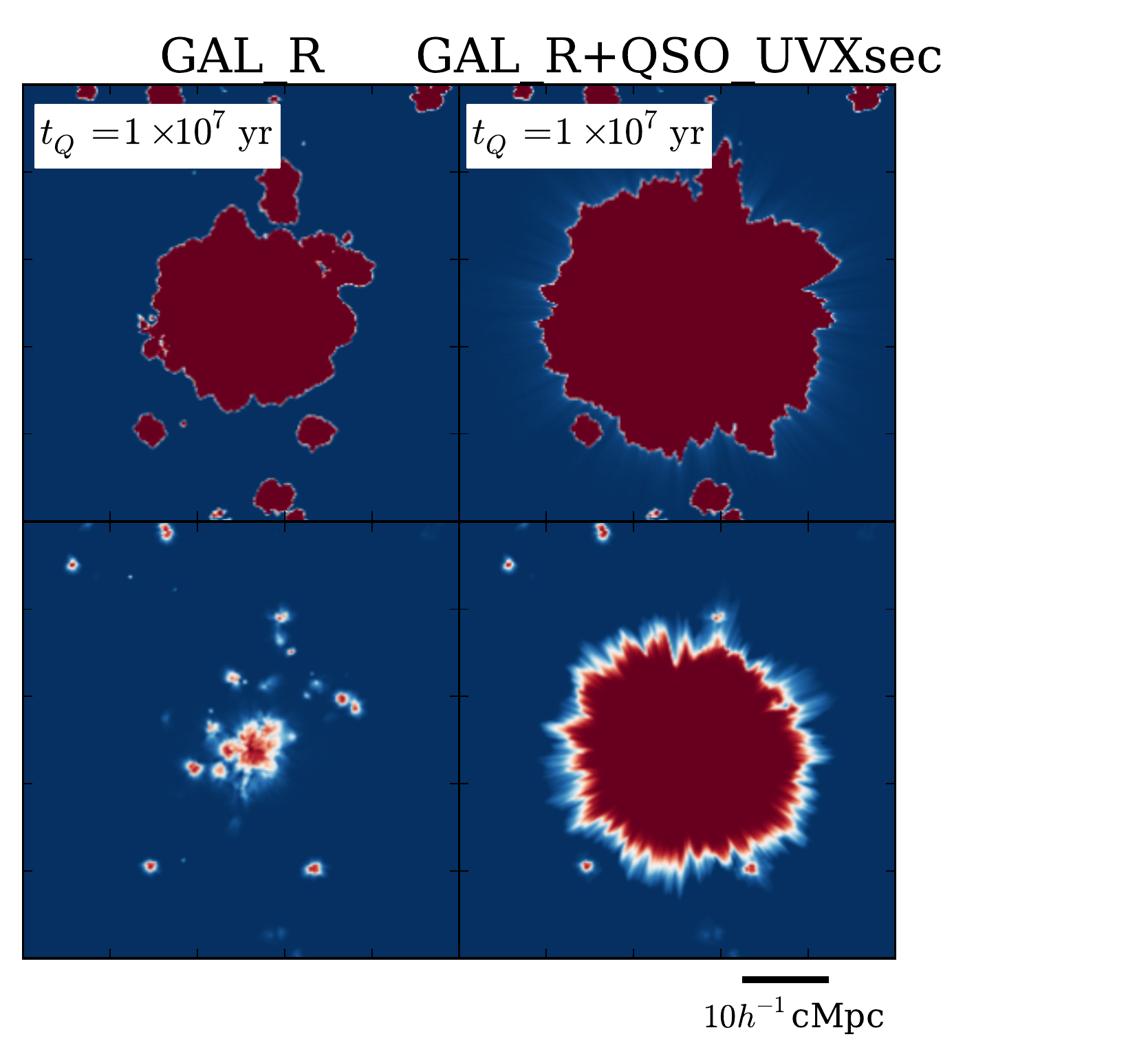}
\hspace{-1.5cm}
  \includegraphics[angle=0,width=\columnwidth]{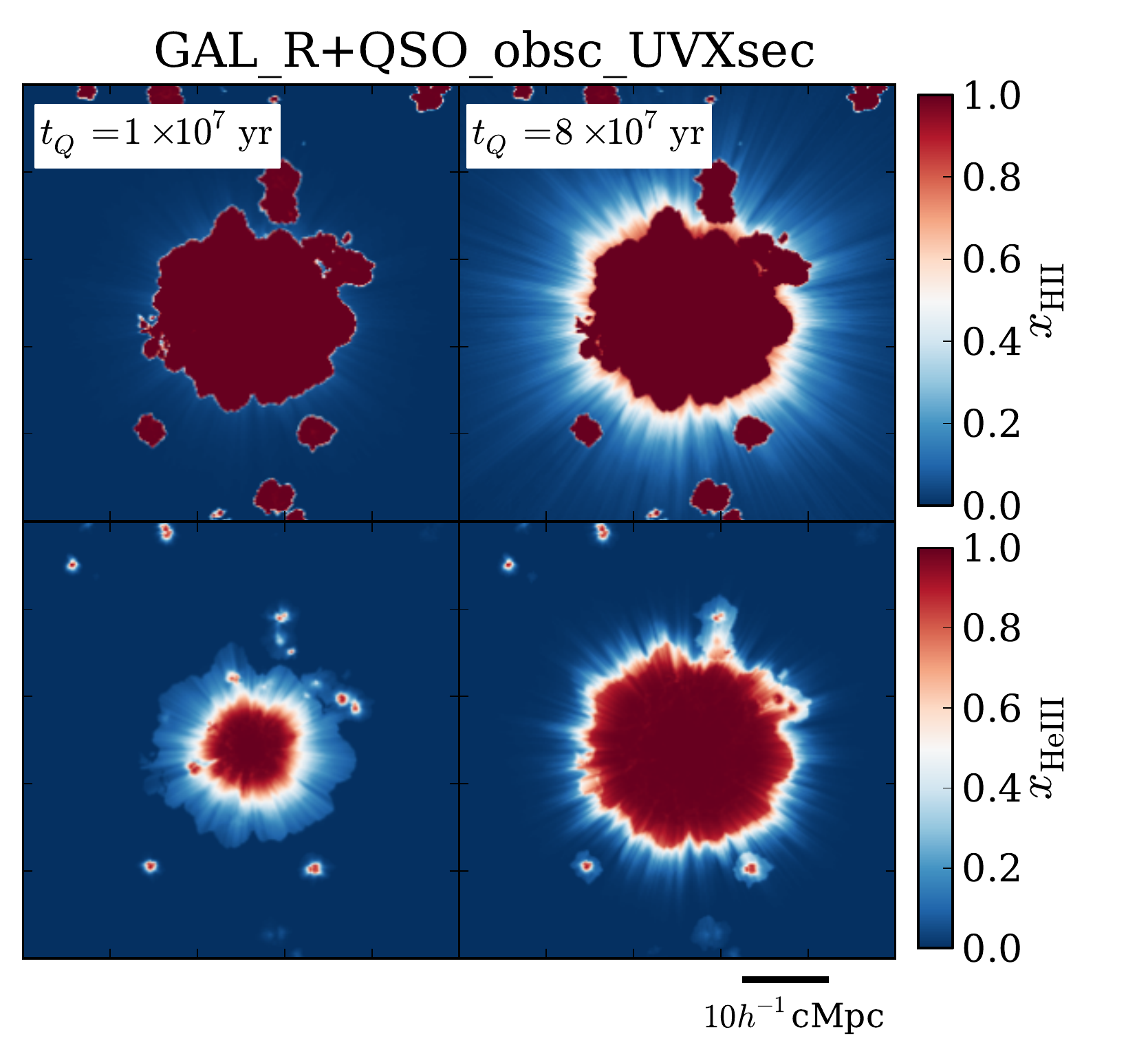}
  \caption[linear-scale maps of $\HII/\HeIII$ fractions for galaxies only, galaxies+QSO, and galaxies+obscured QSO models]{Linear-scale maps of $\HII$ (top panels) and $\HeIII$ (bottom) fractions for galaxies only model GAL\_R, galaxies+QSO model GAL\_R+QSO\_UVXsec, and galaxies+obscured QSO model GAL\_R+QSO\_obsc\_UVXsec at $t_Q=10^7\rm yr$ (first three columns), and  model GAL\_R+QSO\_obsc\_UVXsec at $t_Q=8\times10^7\rm yr$ (last column). The QSO is turned on at $z=10$ in the medium already reionized by galaxies. The length of x- and y-axes is $50h^{-1}\rm cMpc$ and the width of the slice is $195h^{-1}\rm ckpc$. The linear scale was chosen to show structure near the I-front more clearly.}
  \label{fig:obscured_full}
\end{figure*}

\subsection{Effect of an obscured QSO}\label{sec:obscQSO}

In the following, we analyse the RT simulation suite to understand the impact of the QSO's spectral properties on the ionization state of the IGM. 

Here we consider a variant of the QSO model (and galaxies+QSO model) in which all the ionizing UV photons are absorbed by neutral hydrogen  inside the host galaxy (see Figure~\ref{fig:spectra}) and the radiation impacting the surrounding IGM is composed only by the soft X-ray part of the original spectrum. 

Figure \ref{fig:radial_profile_obscured} shows the spherically averaged profiles of the unobscured (QSO\_UVXsec) and obscured (QSO\_obsc\_UVXsec) QSO models. In the obscured QSO model reionization is driven only by soft X-ray photons, and, as a consequence, the $\HII$ and $\HeIII$ regions are much smaller than those from the unobscured QSO.   In addition, the lack of UV photons and the larger mean free path of X-ray photons create I-fronts smoother than those of the unobscured QSO model. We note that the size of the $\HII$ region in the obscured QSO model is consistent with the simple Str\"omgren sphere estimate, which gives $R^{\rm HII}_{\rm I,obscQSO}=R^{\rm HII}_{\rm I,QSO} (\dot{N}^{\rm QSO,obsc}_{ion}/\dot{N}^{\rm QSO}_{ion})^{1/3}\approx3.8h^{-1}\rm cMpc$.

To illustrate the difference in the structure near the $\HII$ and $\HeIII$ I-fronts, Figure \ref{fig:obscured_full} shows the linear-scale maps of $\HII$ and $\HeIII$ fractions for galaxies only model (GAL\_R), galaxies+(unobscured) QSO model (GAL\_R+QSO\_UVXsec), and galaxies+obscured QSO model (GAL\_R+QSO\_obsc\_UVXsec) at the snapshots corresponding to the two QSO lifetimes of $t_Q=10^7\rm~yr$ and $8\times10^7\rm~yr$. 
Unlike the obscured QSO only model, when the contribution from galaxies is included, at $t_Q=10^7\rm~yr$ the $\HII$ I-front structures show little difference between the galaxies+unobscured QSO and galaxies+obscured QSO models. This is because the growth of the I-front is still dominated by galaxies at  $t_Q=10^7\rm~yr$. However, for a longer lifetime, the amount of ionizing photons from the obscured QSO becomes larger than the one from the galaxies, and as a consequence a wider $\HII$ I-front is formed. Thus, the imprint left on the $\HII$ region by an obscured QSO is more distinctive than the one left by an unobscured QSO, even in the presence of galaxies.

The impact of an obscured QSO is more visible in the ionization state of the intergalactic helium. Due to the lack of UV photons and the predominance of soft X-ray photons, the $\HeIII$ I-front produced by an obscured QSO is smoother and wider than the one from an unobscured QSO. 

In summary, an obscured QSO can imprint distinctive features on the surrounding $\HII$ regions by producing a smooth and broad I-front. The effect, though, is noticeable only if the amount of soft X-ray photons exceeds the UV photon budget from the galaxies. 

\section{Thermal state of the IGM}\label{sec:thermal}

In the previous section we examined the ionization state of the IGM. Here we investigate the impact of varying sources, their properties and environments, and the physical processes of radiative transfer on the thermal state of the IGM. 

The thermal structure of the IGM surrounding a QSO is affected in a complex, non-linear way, by the combined input of the galaxies and the QSO. 
We present our results with increasing sophistication to highlight the effect of photoionization heating across the I-fronts (\S~\ref{sec:T_neutral}), pre-heating by X-rays (\S~\ref{sec:T_xray}), and the additional complexity when both galaxies and QSOs are taken into account (\S~\ref{sec:T_reion}).

\subsection{Thermal state induced by galaxies and a QSO in a neutral IGM}\label{sec:T_neutral}

\begin{figure}
  \centering
  \includegraphics[angle=0,width=\columnwidth]{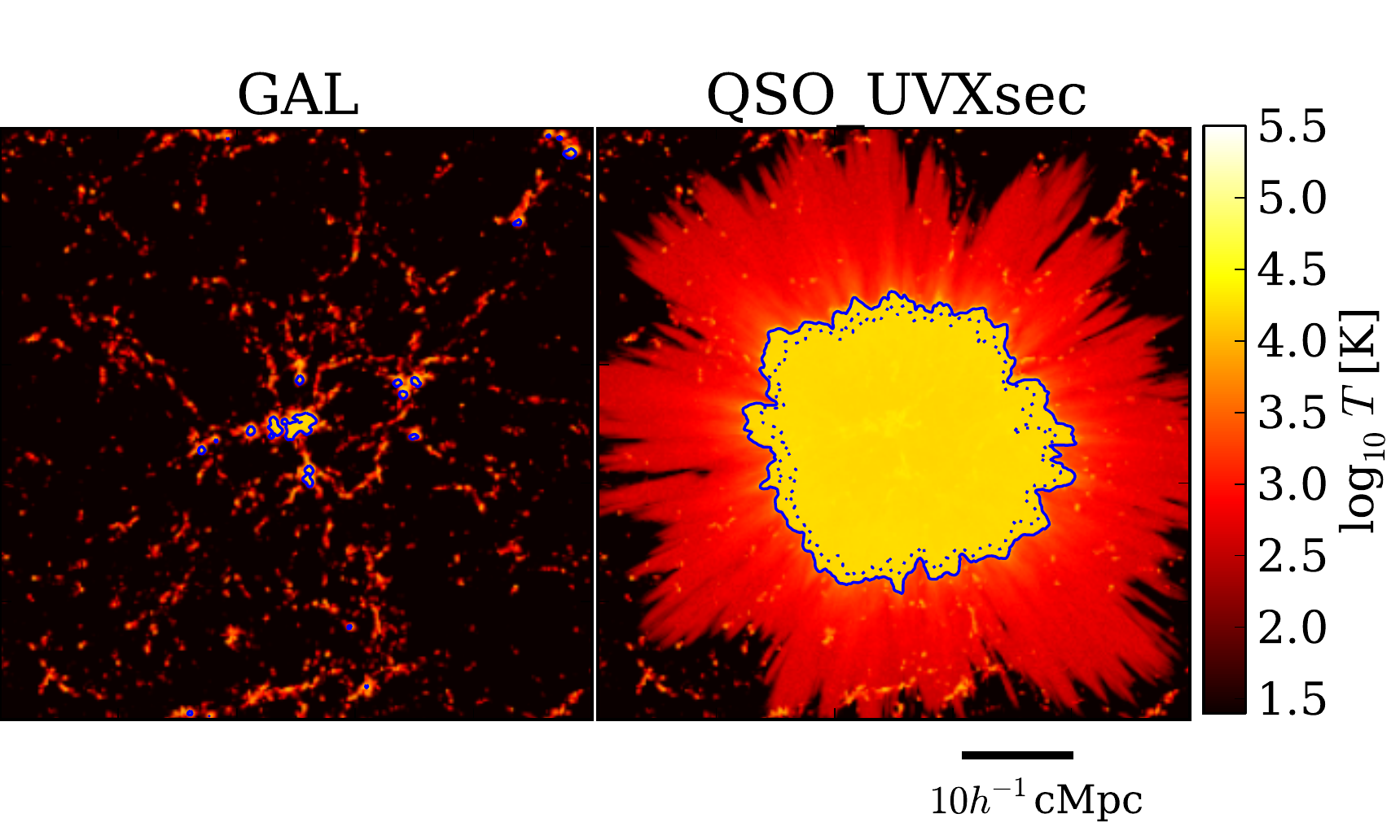}
  \caption[Maps of temperature at $z=10$ for galaxies only and QSO only models]{
Maps of temperature for models GAL (left) and QSO\_UVXsec (right). In both models the sources are turned on in a fully neutral medium at $z=10$ and shine for $10^7\rm~yr$. The maps have a side length of $50h^{-1}$cMpc and the width of the slice is $195h^{-1}\rm ckpc$.
The $\HII$ ($\HeIII$) I-front is shown as blue solid (dotted) contour.}
  \label{fig:T_neutral_run}
\end{figure}

\begin{figure}
  \includegraphics[angle=0,width=\columnwidth]{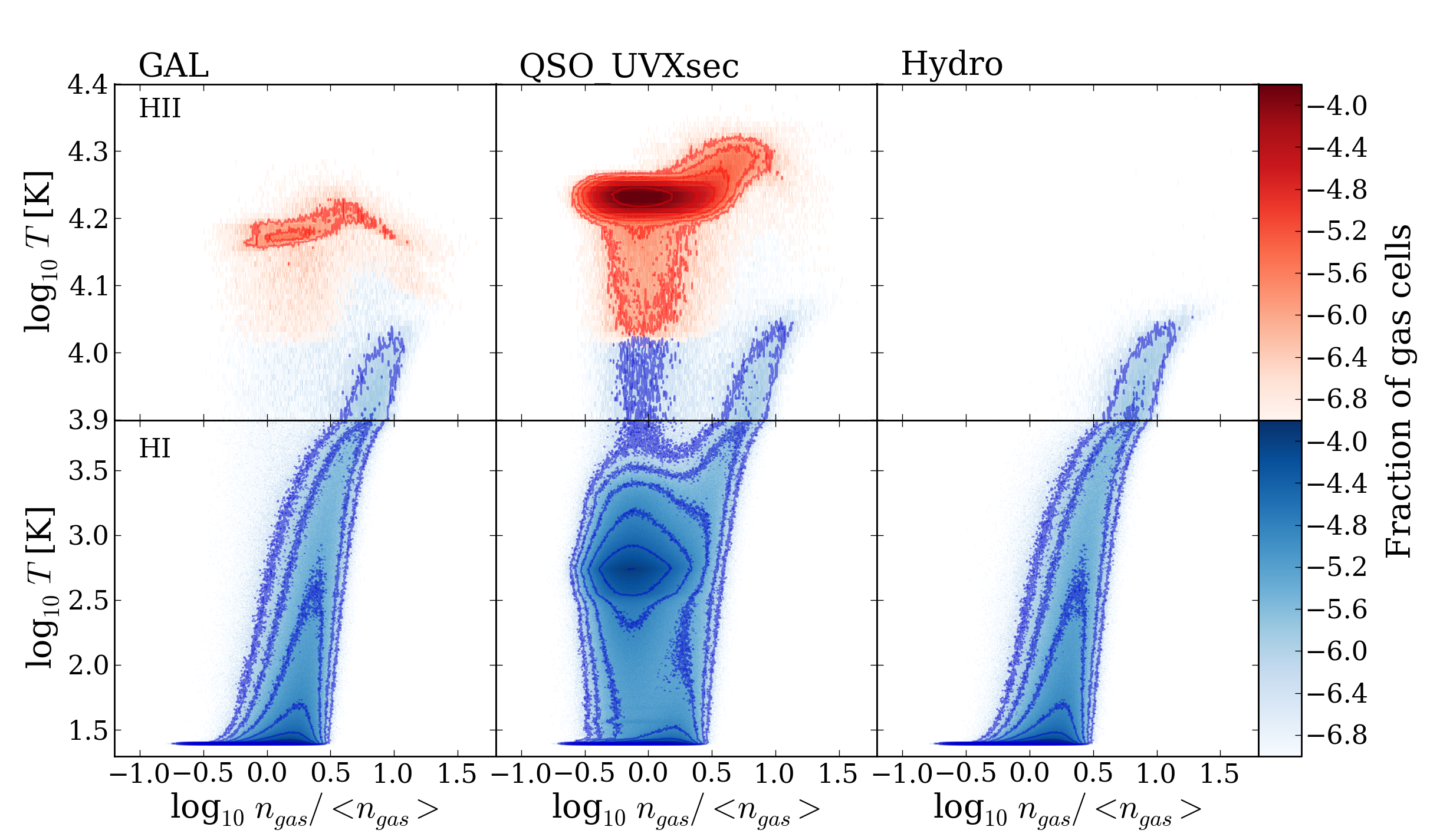}
  \caption[Temperature-density diagram for galaxies only and QSO only models]{Temperature-density diagram for models GAL (left), QSO\_UVXsec (middle), and for the hydrodynamic simulations without RT (right). The sources are turned on in a fully neutral medium at $z=10$ and shine for $10^7\rm~yr$. The red (blue) colour indicates the $\HII$ ($\HI$) regions defined as cells with $\xHII\ge0.5$ $(<0.5)$. The contours are overlaid as a guide. To emphasize the temperature-density relation both in $\HII$ and $\HI$ regions, two different y-axis scales are used. }
  \label{fig:phase_diagram_neutral_run}
\end{figure}

We first present the simplest case where heating is due to photoionization from only  galaxies or QSO, i.e. the equivalent of \S~\ref{neutral_run}. Figure~\ref{fig:T_neutral_run} shows temperature maps for the galaxies (GAL, left) and the QSO (QSO\_UVXsec, right) only model. The contours refer to the position of the $\HII$ (solid) and $\HeIII$ (dotted) I-fronts. In both models, the filamentary thermal structure in the $\HI$ regions is a result of adiabatic compression and hydrodynamic shocks along filaments of cosmic web. Gas within $\HII$ regions is heated by photoionization to $\gtrsim10^4\rm~K$, and therefore the thermal structure is well traced by the position of the I-fronts. 
Similarly to the ionization structure discussed in \S~\ref{neutral_run}, galaxies produce small and patchy heated regions, whereas a large central hot region is present around the QSO, together with warm gas heated by X-rays ahead of the sharp I-front (see \S~\ref{sec:T_xray} for more details).

Figure~\ref{fig:phase_diagram_neutral_run} shows the temperature-density diagram for the two source models discussed above, as well as for the hydrodynamical simulation without RT. The colour indicates $\HII$ ($\HI$) regions where the local $\HII$ fraction is $\xHII\ge0.5(<0.5)$. The temperature of the QSO's $\HII$ region ($\sim17000\rm~K$) is $\sim16\%$ higher than the one of galaxies' $\HII$ regions ($\sim14700\rm~K$), because the QSO harder spectrum heats the gas more effectively per $\HI$ and $\HeI$ ionization (see \S~\ref{sec:photoheating}), and provides an additional heating by $\HeII$ photoionization. 
The temperature of the $\HII$ regions is nearly isothermal in both models, because the gas is ionized within a short period of time ($10^7\rm~yr$) and no substantial adiabatic or recombination cooling (which preferentially cools low  and high density regions, respectively) has taken place. 
While the temperature-density relation of the neutral gas in the galaxies only model is solely determined by hydrodynamic processes (see a comparison between the GAL and Hydro panels), the X-rays photons emitted by the  QSO propagate ahead of the $\HII$ I-front, partially ionizing the gas and heating it to $\sim10^3~\rm K$  (see also \S~\ref{sec:T_xray}).

\subsubsection{Photoionization heating}\label{sec:photoheating}

The extent of the IGM heating due to the passage of the I-fronts depends on the spectral shape of the sources, while it is insensitive to their ionizing photon luminosities. 
The heat gain per ionization across an I-front is (e.g. \citealt{1999ApJ...520L..13A})
\begin{equation}
\langle E_i\rangle=G_i/\Gamma_i\approx\frac{h\nu_i}{2+\alpha_{\rm eff}},
\end{equation}
where $G_i=\int_{\nu_i}^\infty \frac{4\pi J_\nu}{h\nu}\sigma_i(\nu)(h\nu-h\nu_i)d\nu$ is the thermal energy injected by photoionization\footnote{We ignore the impact of secondary ionization for this estimate, although it is included in the simulations.}, $J_\nu$ is the specific intensity, ($h\nu-h\nu_i$) is the excess energy of photoionization above the ionization threshold, $\Gamma_i=\int_{\nu_i}^\infty \frac{4\pi J_\nu}{h\nu}\sigma_i(\nu)d\nu$ is the photoionization rate, and $\alpha_{\rm eff}$ is the effective spectral index at the position of the I-front. The index $i$ indicates the different species, i.e. $i=\HI,\HeI,\HeII$.
In the optical thin limit $\alpha_{\rm eff}=\alpha_G$ or $\alpha_{\rm eff}=\alpha_Q$, while when spectral hardening occurs the effective spectral index becomes smaller, i.e. $\alpha_{\rm eff}<\alpha_G$ or $\alpha_{\rm eff}<\alpha_Q$.

From the energy conservation across the I-front, we obtain $(3/2)k_Bn_{gas}^{\rm after}\Delta T_i\approx n_i^{\rm before}\langle E_i\rangle$, where $\Delta T_i$ is the temperature jump across the I-front, and $n_{gas}^{\rm after}$ ($n_i^{\rm before}$) is the number density of gas of $i$-th species after (before) the passage of the I-front.  

The total temperature jumps ($h\nu\rightarrow\infty$) across the $\HII$, $\HeII$, and $\HeIII$ I-fronts are then given by\footnote{For this estimate, we have approximated the photoionization cross sections of $\HI$, $\HeI$, and $\HeII$ as power-laws, $\sigma_{\rm HI}\approx6.3\times10^{-18}(\nu/\nu_{\rm HI})^{-3}{~\rm cm^2},~\sigma_{\rm HeI}\approx7.8\times10^{-18}(\nu/\nu_{\rm HeI})^{-2}{~\rm cm^2},\sigma_{\rm HeII}\approx1.6\times10^{-18}(\nu/\nu_{\rm HeII})^{-3}{~\rm cm^2}$, respectively (\citealt{2011piim.book.....D}).}
\begin{align}
&\Delta T_{\rm HII}=\frac{2}{3k_B}\frac{\langle E_{\rm HI}\rangle}{2+Y/X}\approx5.0\times10^4(2+\alpha_{\rm eff})^{-1}\rm~K,\\
&\Delta T_{\rm HeII}=\frac{2}{3k_B}\frac{\langle E_{\rm HeI}\rangle}{2(X/Y)+2}\approx7.6\times10^3(1+\alpha_{\rm eff})^{-1}\rm~K,\\
&\Delta T_{\rm HeIII}=\frac{2}{3k_B}\frac{\langle E_{\rm HeII}\rangle}{2(X/Y)+3}\approx1.6\times10^4(2+\alpha_{\rm eff})^{-1}\rm~K.
\end{align}
Thus the temperature jump after the passage of all the I-fronts is $\approx2.2\times10^4\rm~K$ for a QSO-type spectrum with $\alpha_{\rm eff}=1.5$, whereas $\approx1.5\times10^4(1.2\times10^4)\rm~K$ for a galaxy-type spectrum with $\alpha_{\rm eff}=3$ (without $\HeII$ reionization). In this estimate, a QSO-type spectrum is expected to heat up the gas $\sim 80\%$ more effectively than a galaxy-type spectrum. This is somewhat larger than the difference found in our simulations. As shown in \S~\ref{sec:T_xray}, neglecting secondary ionization, the temperature inside the QSO $\HII$ region becomes $\sim20000\rm~K$. In addition, the cooling is neglected in the simplified argument. Taking into account these two factors, the above simple estimate explains, as a first approximation, the difference in the temperature seen in the simulations.

The above simplified argument demonstrates the physical origin of the temperature jump across the $\HII$, $\HeII$, and $\HeIII$ I-fronts, i.e. the energy conservation of spectral-weighted excess energy of photo-electrons across the I-fronts and the change of gas particle number.  A harder spectrum more efficiently heats up the gas per ionization because the fraction of higher energy photons relative to photons near the ionization threshold is larger for a smaller spectral index.

\begin{figure}
\centering
  \includegraphics[angle=0,width=\columnwidth]{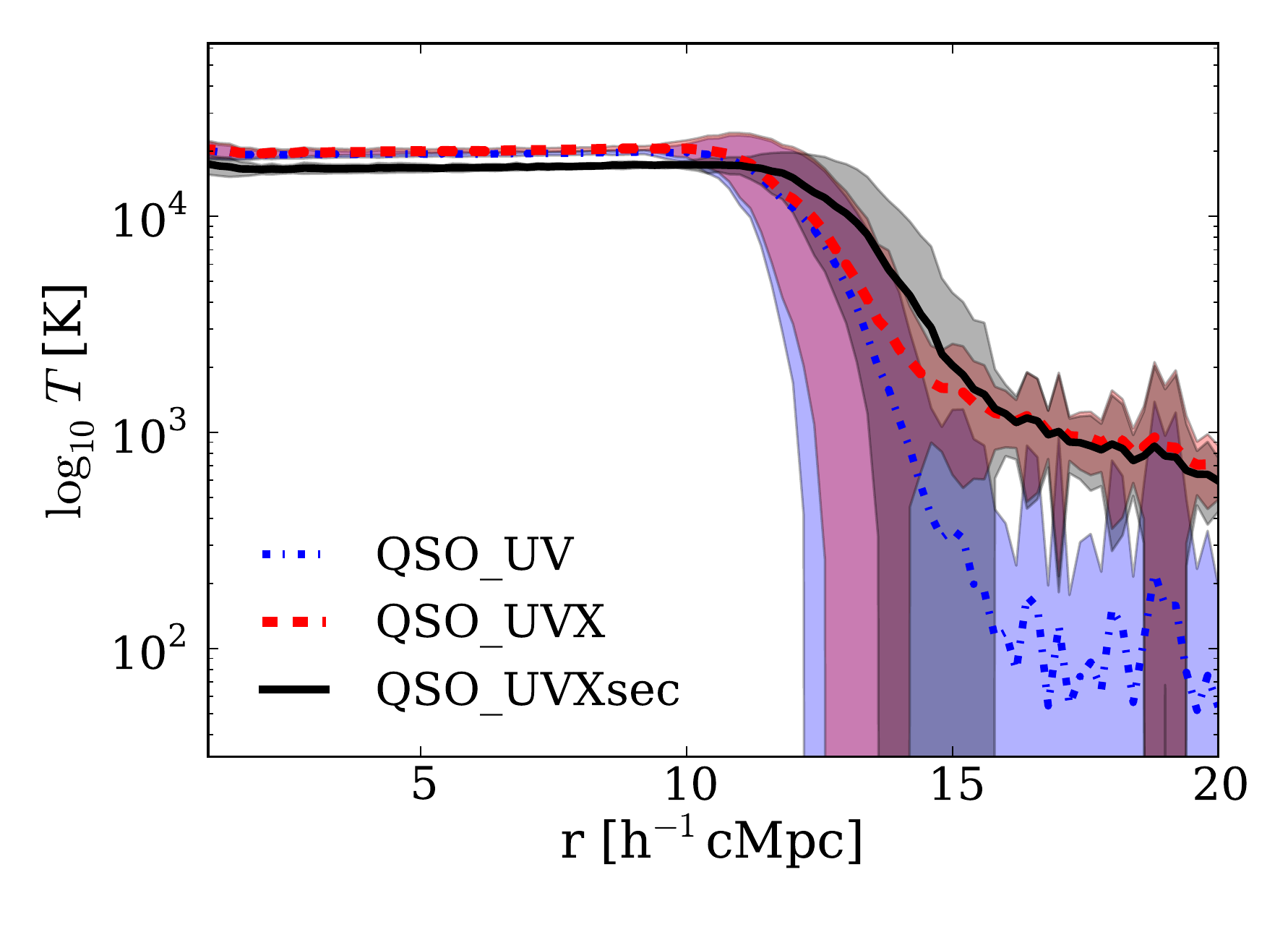}
  \caption[Spherically averaged temperature profiles for QSO models with various X-ray physics]{Spherically averaged temperature profiles around a QSO including only UV photons (QSO\_UV, blue dotted), UV and X-ray photons (QSO\_UVX, red dashed), UV and X-ray photon and secondary ionization (QSO\_UVXsec, black solid). The shaded regions indicate the $1\sigma$ scatter around the mean.}
  \label{fig:radial_profile_xray}
\end{figure}

\begin{figure*}
  \centering
  \includegraphics[angle=0,width=\textwidth]{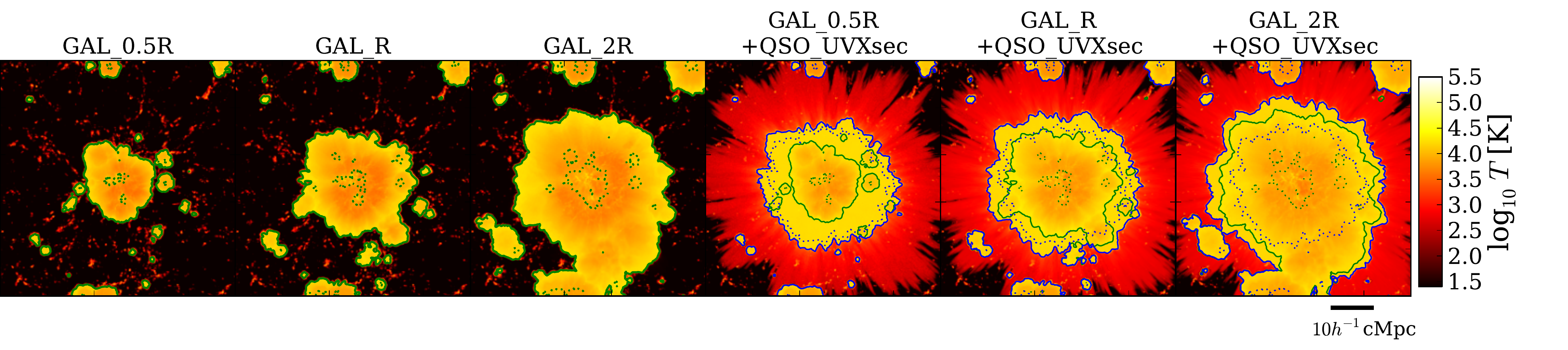}
  \caption[Maps of temperature for galaxies only and galaxies+QSO models]{Maps of temperature at $z=10$ for galaxies only models GAL\_0.5R, GAL\_R, GAL\_2R, and galaxies+QSO models GAL\_0.5R+QSO\_UVXsec, GAL\_R+QSO\_UVXsec, GAL\_2R+QSO\_UVXsec. In both sets of models galaxies start to reionize the IGM at $z=15$, while the QSO is turned on at $z=10$ for a duration of $10^7\rm~yr$ in the medium already reionized by the pre-existing galaxies. The maps have a side length of $50h^{-1}\rm cMpc$ and the width of the slice is $195h^{-1}\rm ckpc$. The galaxies' and QSO's $\HII$ ($\HeIII$) I-front is shown as green and blue solid (dotted) contour, respectively. In the three rightmost panels, the $\HII$ I-front of the pre-existing galaxies are overlaid to aid the interpretation of the figure (see text).}
  \label{fig:T_full_run}
\end{figure*}

\subsection{Effect of X-rays and secondary ionization}\label{sec:T_xray}

Figure~\ref{fig:radial_profile_xray} shows the spherically averaged profiles of temperature for three QSO only models including only UV photons (QSO\_UV), UV and X-ray photons neglecting secondary ionization (QSO\_UVX), and including it (QSO\_UVXsec).
Compared to a UV only case in which the temperature beyond the I-front drops to $\sim 100$~K, X-ray photons heat the gas to $\sim10^3\rm~K$. Including secondary ionization lowers the temperature by $\sim10\%$ because some of the excess energy of fast photo-electrons is diverted from heating to collisional ionization of hydrogen and helium.
This effect is clearly visible also in the fully ionized region, at $<10h^{-1}\rm~cMpc$, where the temperature is reduced by $\sim 20\%$, from $\sim20000\rm~K$ to $\sim17000\rm~K$. Photoheating alone by X-ray photons, though, does not affect the temperature in the $\HII$ region, as it is set by the UV photons (as shown in \S~\ref{sec:photoheating}). 

Near the I-fronts at $\sim10-15h^{-1}\rm cMpc$, the QSO model with secondary ionization shows a temperature higher than the model without it. This is related to the fact that our secondary ionization model pushes the I-front slightly outwards, i.e. the location of the photoionization heating by UV photons at the I-front is displaced. Therefore, secondary ionization indirectly increases the radius of the photoheated region.
  
In summary, X-ray photons heat the gas beyond the I-front because of their larger mean free path, but the inclusion of secondary ionization lowers the photoheating efficiency, because some of the excess energy of fast photo-electrons is diverted to collisional ionization of hydrogen and helium.

\subsection{Thermal state induced by galaxies and a QSO in a pre-ionized IGM}\label{sec:T_reion}

If a QSO turns on in a medium ionized by pre-existing galaxies, the thermal state of the IGM is determined by the combined effect of both galaxies and the central QSO. 

Figure~\ref{fig:T_full_run} shows temperature maps for galaxies only (GAL\_0.5R, GAL\_R, GAL\_2R) and galaxies+QSO (GAL\_0.5R+QSO\_UVXsec, GAL\_R+QSO\_UVXsec, and GAL\_2R+QSO\_UVXsec) models, in which the QSO is turned on in a medium previously ionized by pre-existing galaxies. The green and blue solid (dotted) contours indicate the location of $\HII$ ($\HeIII$) I-fronts due to galaxies and a QSO. As already noticed in Figure \ref{fig:T_neutral_run}, the thermal structure is traced by the position of the $\HII$ I-front due to the photoionization heating of hydrogen. Differently from a case in which reionization is driven by a QSO only (\S~\ref{sec:T_neutral}), pre-existing galaxies place the $\HII$ I-fronts further ahead of the QSO $\HeIII$ I-fronts. Thus, a large fraction of the thermal structure inside the $\HII$ region is determined by the pre-existing galaxies instead of the central QSO. The effect is more obvious as the total ionizing photon emissivity from galaxies increases. Similarly to what already noticed for the $\HII$ regions (\S~\ref{reion_run}), except for the pre-heating by X-rays from a QSO ahead of the $\HII$ I-front, the morphology and extent of the gas heated at $\sim 10^4\rm~K$ is very similar in both galaxies only and galaxies+QSO models, as it is mainly determined by the UV photons emitted by galaxies.

In the following sections, we examine the thermal structure in more detail, focusing on the effect of galaxies and the additional heating by a QSO.

\subsubsection{Heating by galaxies only}\label{sec:galaxy_heating}

A galaxy-type spectrum contributes to the thermal structure of the IGM primarily by $\HI$ and $\HeI$ photoionization heating.  Figure~\ref{fig:T_radial_full} shows the spherically averaged temperature profiles. Furthermore, in Figure~\ref{fig:phase_diagram_full_run} the temperature-density diagram of $\HII$ (red) and $\HI$ (blue) regions shows that in galaxies only models (left three panels) the maximum temperature ($\sim2\times10^4\rm~K$) depends weakly on the total ionizing photon emissivity, whereas the fraction of gas heated close to the maximum temperature increases with total emissivity. This is because, for a fixed source spectrum, the heating rate per ionization is fixed (\S~\ref{sec:photoheating}), and thus the heating efficiency remains the same regardless of the total ionizing photon production rate.
When galaxies ionize the IGM gradually rather than in a short period of time (as discussed in \S~\ref{sec:T_neutral}), the $\HII$ temperature distribution is no longer isothermal as the gas engulfed earlier by the I-front has more time to cool. Furthermore, as the cooling time depends on the gas density, it is expected to vary according to location. 
This is clearly visible in the top panel of Figure~\ref{fig:T_radial_full}, where the spherically averaged temperature profiles of galaxies only models is plotted. Here the temperature in the inner part of the $\HII$ region is lower than the gas temperature immediately behind the I-front\footnote{Note that because photoionization of the residual neutral gas inside the $\HII$ region can contribute to heating, the temperature toward the inner parts does not decrease monotonically.}, and the spread is large (see the difference in the spread in the temperature-density relation inside $\HII$ regions shown in Figures~\ref{fig:phase_diagram_neutral_run} and \ref{fig:phase_diagram_full_run}).

On the other hand, the thermal structure in the neutral regions is nearly identical in all galaxies only models, as, similarly to \S~\ref{sec:T_neutral}, hydrodynamical processes, and not UV photons, are responsible for controlling the temperature in the neutral gas.

\subsubsection{Heating by galaxies and a QSO}\label{sec:qso_heating}

\begin{figure}
  \centering
  \includegraphics[angle=0,width=\columnwidth]{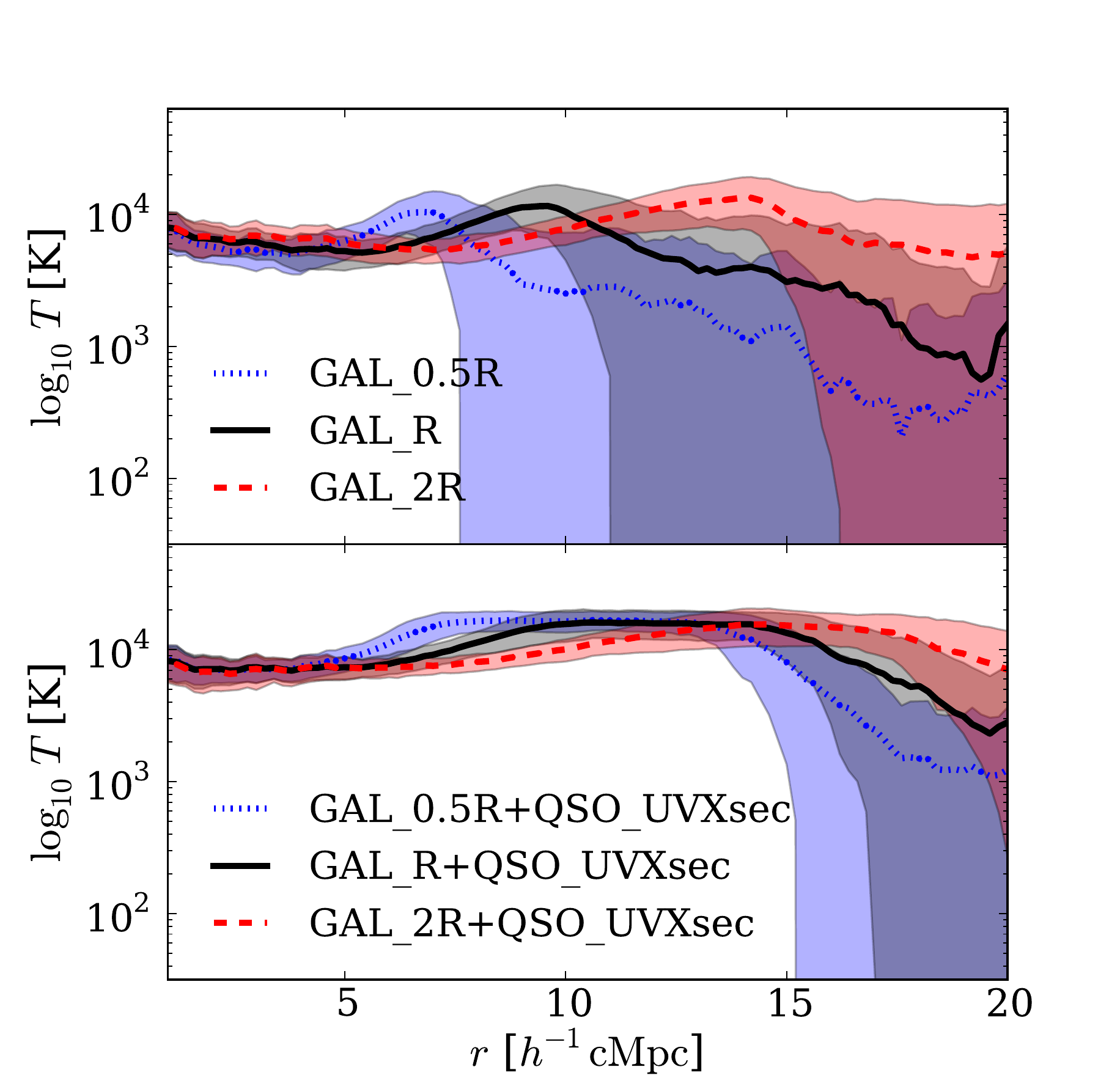}
  \caption[Spherically averaged temperature profiles of galaxies only and galaxies+QSO models]{Spherically averaged temperature profiles of galaxies only models (top panel) GAL\_0.5R (blue dotted line), GAL\_R (black solid), GAL\_2R (red dashed), and galaxies+QSO models (bottom panel) GAL\_0.5R+QSO\_UVXsec (blue dotted), GAL\_R+QSO\_UVXsec (black solid), GAL\_2R+QSO\_UVXsec (red dashed). The shaded regions indicate the $1\sigma$ scatter around the mean.}
  \label{fig:T_radial_full}
\end{figure}

\begin{figure*}
  \includegraphics[angle=0,width=\textwidth]{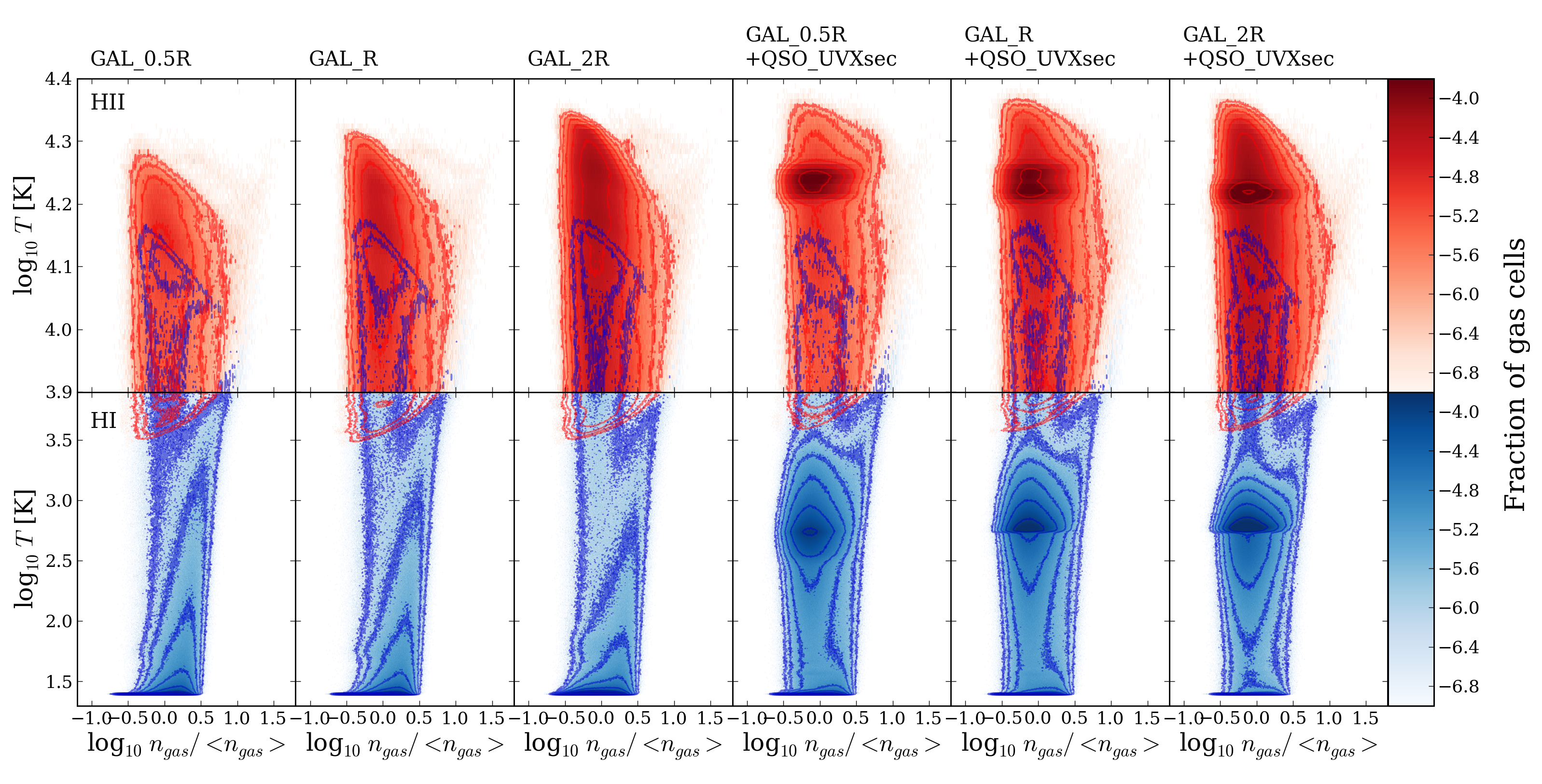}
  \caption[Temperature-density diagram for galaxies only and galaxies+QSO models]{Temperature-density diagram for galaxies only models GAL\_0.5R, GAL\_R, GAL\_2R and galaxies+QSO models GAL\_0.5R+QSO\_UVXsec, GAL\_R+QSO\_UVXsec, GAL\_2R+QSO\_UVXsec. The red (blue) colour indicates the $\HII$ ($\HI$) regions defined as cells with $\xHII\ge0.5$ $(<0.5)$. The contours are overlaid as a guide. To emphasize the temperature-density relation both in $\HII$ and $\HI$ regions, two different y-axis scales are used. 
}
  \label{fig:phase_diagram_full_run}
\end{figure*}

Figure~\ref{fig:T_full_run} shows that as the QSO increases the extent of the fully ionized region, the radius at which the gas is heated to $\sim10^4\rm~K$ is larger than in the corresponding galaxies only models. However, both the average profiles (Figure~\ref{fig:T_radial_full}) and the overall temperature-density relations (Figure~\ref{fig:phase_diagram_full_run}) are similar to those of the corresponding galaxies only models. This indicates that a QSO plays a small role in controlling the thermal state in the inner parts of the $\HII$ region, where the pre-existing galaxies have ionized the IGM before the QSO activity.    

On the other hand,  the top panels in Figure~\ref{fig:phase_diagram_full_run} show that galaxies+QSO models reach temperatures which are typically higher than those in the corresponding galaxies only models. The differences become less evident with increasing ionizing emissivity from the galaxies (from left to right in the three rightmost panels).
The reason for this is that when both the $\HII$ and $\HeIII$ I-fronts from the QSO are larger than the pre-existing $\HII$ region due to galaxies, the gas experiences the highest heating. However, when only the $\HII$ I-front from the QSO is larger than the pre-existing $\HII$ region (but the QSO $\HeIII$ I-front is still inside it), the portion of the gas `newly' heated up by $\HI$ photoionization of a QSO has not yet undergone QSO-driven $\HeII$ photoionization. This lack of $\HeII$ photoionization by the QSO results in a slightly lower temperature. This non-linear heating process, which depends on the relative position of the various I-fronts, explains why QSO heating becomes less efficient with increasing contribution from galaxies.

\begin{figure}
  \centering
  \includegraphics[angle=0,width=\columnwidth]{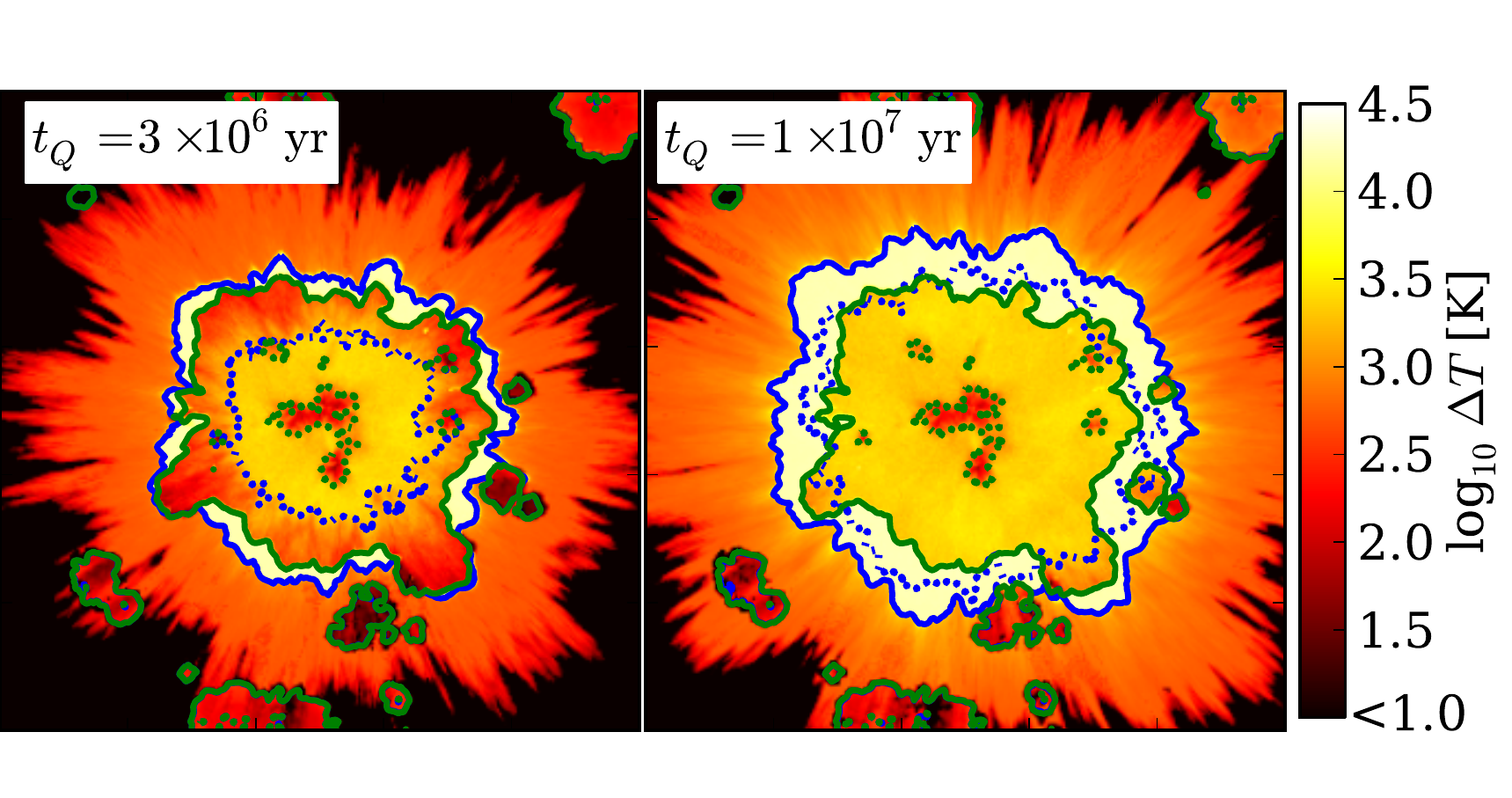}
  \caption[Differential temperature map for QSO heating]{Differential temperature maps between models GAL\_R+QSO\_UVXsec and GAL\_R for QSO lifetimes of $t_Q=3\times10^6\rm~yr$ (left panel) and $t_Q=10^7\rm~yr$ (right). The blue solid (dotted) contours indicate the location of the QSO's $\HII$ ($\HeIII$) I-front, and the green solid (dotted) contours indicate the location of the $\HII$ ($\HeIII$) I-front created by the pre-existing galaxies.}
  \label{fig:dT}
\end{figure}

\begin{figure}
  \centering
  \includegraphics[angle=0,width=0.9\columnwidth]{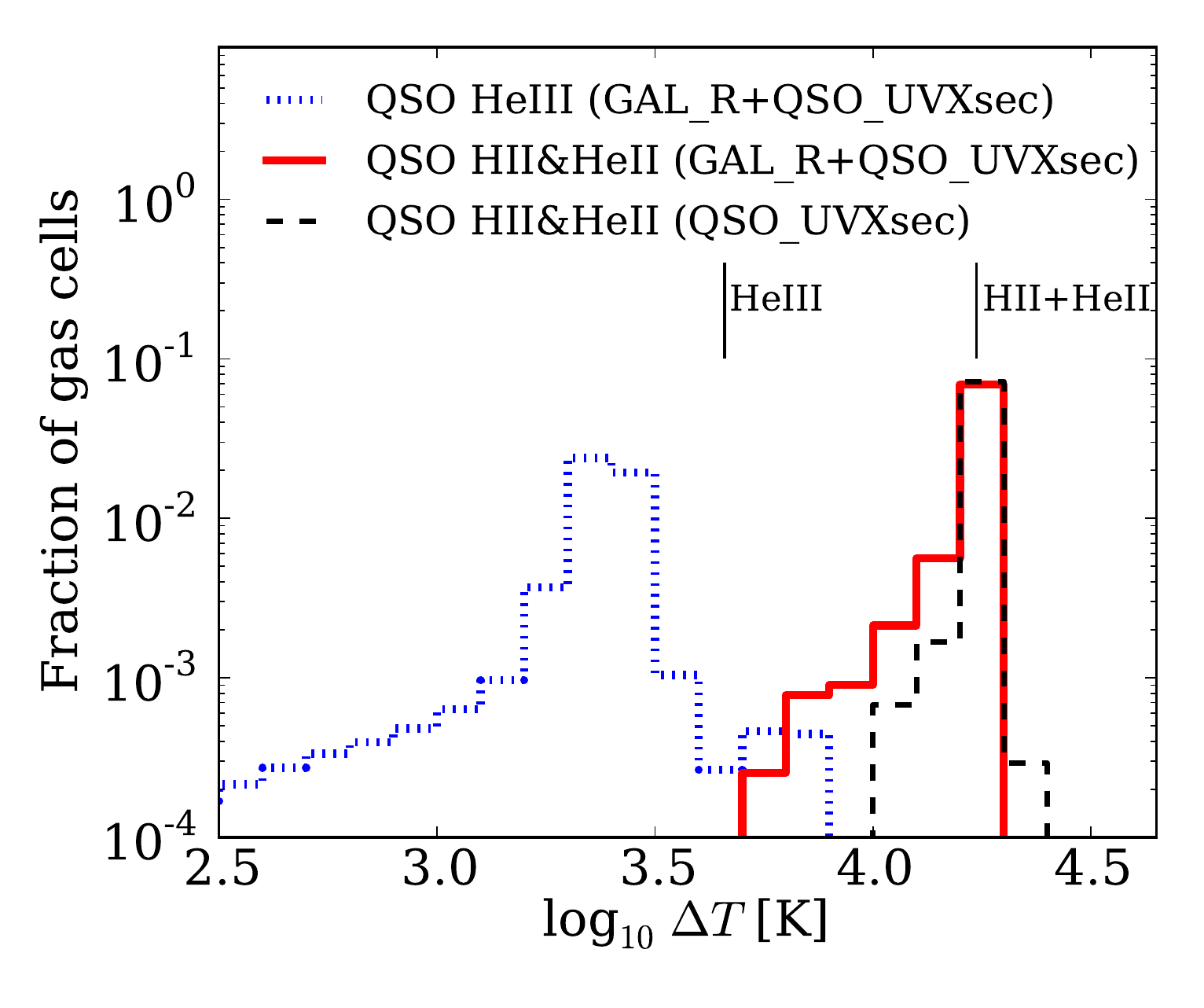}
  \caption[Differential temperature map for QSO heating]{Fraction of cells heated by $\Delta T$ due to the QSO in GAL\_R+QSO\_UVXsec and QSO\_UVXsec models at $t_Q=10^7\rm~yr$. The red solid (blue dotted) histogram refers to the distribution of the gas that has been swept by the $\HII+\HeII$ ($\HeIII$) I-front. For a comparison, the black dashed histogram shows the distribution of the gas that has been swept by the $\HII+\HeII$ I-front in the QSO only model QSO\_UVXsec. The two vertical lines are the analytic estimates given in \S~\ref{sec:photoheating}.} 
\label{fig:histdT}
\end{figure}

This can be more clearly seen in 
Figure~\ref{fig:dT}, which shows differential maps of temperature between GAL\_R+QSO\_UVXsec and GAL\_R at two different QSO lifetimes. The solid (dotted) blue and green contours indicate the location of $\HII$ ($\HeIII$) I-fronts produced by the QSO and the galaxies, respectively. The heating provided by the QSO in addition to the one by pre-existing galaxies is clearly associated with the positions of the various I-fronts. At $t_Q=10^7\rm~yr$ the gas newly swept by the QSO $\HII$ I-front (between the blue solid and green solid contours) receives the highest temperature boost ($\Delta T\sim17000\rm~K$, see Fig.~\ref{fig:histdT}), while the gas newly swept by the QSO $\HeIII$ I-front (between the blue dashed and green dashed contours) is additionally heated by $\Delta T\sim3000\rm~K$ (see Fig.~\ref{fig:histdT}). Note that at $t_Q=3\times10^6\rm~yr$ the radius of the QSO's $\HeIII$ region has not yet exceeded the radius of the $\HII$ region produced by the pre-exisiting galaxies, so the gas between these two radii experiences a heating from the QSO of only $\Delta T\sim300\rm~K$.  At $t_Q=10^7\rm~yr$ the QSO $\HeIII$ region has grown large enough to heat up all the $\HII$ gas already heated by the pre-existing galaxies. It is worth highlighting that, when a QSO shines in a region already reionized by galaxies, the temperature increment is mainly driven by $\HeII$ reionization, whereas outside the pre-existing $\HII$ region from galaxies the main contribution comes from $\HI$ reionization by the central QSO.
In general, the QSO heating is more effective where its radiation reaches the neutral gas, while photoionization of residual neutral gas in regions already ionized by pre-existing sources does not contribute substantially.
Additionally, because of its harder spectrum, a QSO can heat gas more effectively than galaxies (by about a factor of $\Delta T^{\rm QSO}/\Delta T^{\rm GAL}\approx [2+\alpha_{\rm GAL}]/[2+\alpha_{\rm QSO}]\approx1.4$). As a result,
the gas swept {\it only} by the QSO's I-fronts can reach temperatures higher than the gas which has been pre-ionized by galaxies.

For a more quantitative analysis, Figure~\ref{fig:histdT} shows the fraction of gas heated by the passage of the QSO's $\HII$ and $\HeIII$ I-fronts. Because $\HeI$ and $\HI$ reionization occur almost simultanously (see \S~\ref{reion_run}), the position of the $\HeII$ I-front is identical to that of the $\HII$ I-front. Their passage increases the temperature by $\Delta T\sim17000\rm~K$. This is well captured by the analytic estimates in \S~\ref{sec:photoheating}. The fraction of gas heated by the QSO's I front in the galaxies+QSO and QSO only models is very similar. This is not surprising because the luminosity of the QSO is identical in the two models, and thus it can ionize and photoheat the same volume of gas, although with different geometry. On the other hand, the QSO's $\HeIII$ I-front increases the temperature by $\sim3000\rm~K$. Although this is somewhat lower than the simple analytic estimate, the results support the physical picture that the passage of the QSO's $\HeIII$ I-front is a substantial source of heating inside the pre-existing $\HII$ region.

In summary, the hydrogen and helium reionization driven by galaxies and QSOs has a non-linear impact on the gas thermal structure, which differs substantially from the one obtained in a simpler reionization model driven by galaxies or QSOs only. The relative position of the $\HII$ I-front of galaxies and the $\HII/\HeIII$ I-fronts of the QSO determines the thermal structure of the IGM. In general, higher temperatures can be achieved if the gas is heated {\it only} by QSOs, while when the gas has been ionized by pre-existing galaxies the QSO heating is less efficient.
The order of the $\HII$ and $\HeII/\HeIII$ I-fronts is discussed in Appendix \ref{sec:I-front_order} for interested readers.

\section{Does the QSO imprint unique features on HII regions?}\label{sec:comparison}

One of the outstanding questions regarding the role of QSOs during reionization is whether they leave distinctive features in the structure of $\HII$ regions.
The answer to this depends on ({\it i}) the relative amount of the ionizing photons emitted by a QSO and galaxies since the onset of reionization, and ({\it ii}) the spectrum of the QSO. 

As a QSO is likely born in a region overdense in galaxies (e.g. \citealt{2005ApJ...620...31Y,2007ApJ...670...39L,2007MNRAS.380L..30A,2008MNRAS.386.1683G}), while the QSO dominates the local photon emissivity during its activity, galaxies contribute to the ionizing budget for a longer time, and thus the structure of $\HII$ regions is determined by the relative importance of the two contributions (see discussion in \S~\ref{sec:clustering}, but also \citealt{2007ApJ...670...39L} and \citealt{2012MNRAS.424..762D}). 
When the total ionizing budget from a QSO dominates over that of the surrounding galaxies since the onset of reionization, the local environment is more strongly affected by the QSO and the resulting $\HII$ region is more spherical than those obtained when galaxies are more relevant. This limit corresponds to the scenarios considered by \citet{2008MNRAS.384.1080T} and \citet{2013MNRAS.429.1554F}, and our models discussed in \S~\ref{neutral_run}. 
A consequence of this argument is that a QSO can more easily imprint a distinctive $\HII$ region at earlier than later times because of a smaller cumulative contribution from galaxies (\citealt{2008MNRAS.386.1683G}).
On the other hand, in \citet{2007ApJ...670...39L}, \citet{2012MNRAS.424..762D}, and our galaxies+QSO models, the contribution from galaxies is more important. 
In this case, a QSO does not change drastically the morphology of the $\HII$ region, although it enhances its sphericity by a degree increasing with the ionizing photon production rate of the QSO. 

The sphericity of the $\HII$ region is, however, broken when ({\it i}) the QSO I-front merges with $\HII$ regions of galaxies (e.g. \citealt{2013MNRAS.429.1554F}), ({\it ii}) the QSO I-front encounters high density neutral gas, e.g. Lyman-limit systems, which casts shadows (see \S~\ref{neutral_run}, \ref{reion_run} and \citealt{2015MNRAS.454..681K}), and ({\it iii}) a QSO shines anisotropically. In other words, the degree of sphericity is influenced by both the QSO-galaxy and QSO-absorber clustering (\citealt{2015MNRAS.454..681K}), and by a beamed emisson, which might serve as another indicator of the QSOs presence.

The spectrum of a QSO plays a pivotal role in determining the morphology of $\HII$ regions. By studying QSO models with UV-obscured and unobscured spectra, previous works (\citealt{2005MNRAS.360L..64Z}; \citealt{2008MNRAS.385.1561K}; \citealt{2008MNRAS.384.1080T}) concluded that an obscured QSO spectrum produces a wide I-front. Thus, as a QSO spectrum becomes more UV-obscured and harder, it can imprint a distinctive ionization structure, providing smoother and more extended I-fronts, and more spherical $\HII$ regions. On the other hand, an unobscured QSO emitting UV photons renders the I-front more similar to one by galaxies. 
Our results are consistent with this picture, but add another ingredient. Although for a single LOS, a harder and UV-obscured spectrum thickens the I-front, large fluctuations in the I-front position are observed in many LOSs. Therefore, in order for the thickness of the I-front to be used as an indicator of the source type, it must exceed the scatter observed along the LOSs. In addition, the soft X-ray photon budget from the QSO must dominate the UV photons contribution from pre-existing galaxies in order for an obscured QSO to imprint a distinctive morphology on the $\HII$ region. 

In summary, a QSO leaves a distinctive imprint on an ionized region in terms of its sphericity and the thickness of the I-front when the QSO has a hard UV-obscured spectrum and/or when its contribution to the total ionizing photon budget in its local environment dominates the one from galaxies since the onset of reionization. If instead the QSO spectrum is not obscured and the ionizing photon budget of the local galaxies is comparable to or larger than the one of the QSO, no obvious features would appear in the $\HII$ region to indicate the presence of a QSO. In conclusion, a large quasi-spherical $\HII$ region at high redshift is not necessarily associated with the presence of a QSO.

\section{Observational implications}\label{sec:obs}

A natural question is whether observations can be envisioned in order to provide information about the role of galaxies and QSOs in driving reionization. 

Tomography of the 21 cm line from neutral hydrogen can directly image the morphology of $\HII$ regions during reionization (see the reviews by \citealt{2006PhR...433..181F,2010ARA&A..48..127M,2013ExA....36..235M}). Although the structure of a QSO $\HII$ region should appear as a coldspot in a 21 cm image and offer information about the ionizing sources (\citealt{1997ApJ...475..429M,2000ApJ...528..597T,2004ApJ...610..117W,2008MNRAS.384.1080T,2008MNRAS.385.1561K,2012MNRAS.424..762D}), as emphasized in this work, the morphology of $\HII$ regions in high-$z$ QSO environments is complicated by the contribution of the surrounding galaxies and the spectral shape of the QSO. This implies that some degeneracy will remain when trying to constrain the properties of ionizing sources using 21 cm data alone. We note that the appearence of $\HII$ regions as coldspots depends on the surrounding thermal structure. We investigate this topic more in detail in a forthcoming paper.

Galaxy surveys in the direction of coldspots in 21 cm images can provide important complementary information (\citealt{2015ApJ...800..128B}). This strategy can be realized, for example, in ELAIS-N1 field targeted by LOFAR (\citealt{2014A&A...568A.101J}) and Subaru/Hyper-SuprimeCam Survey\footnote{\url{http://www.naoj.org/Projects/HSC/surveyplan.html}}. Alternatively, 21 cm images can be formed in the direction of high redshift QSOs after they are found by galaxy surveys. Ongoing wide field surveys such as Dark Energy Survey and Subaru/Hyper-SuprimeCam Survey are expected to provide $\sim10$ new QSOs at $z\sim7$ (\citealt{2015MNRAS.454.3952R,2016arXiv160302281M}). This approach would allow us to directly infer the expected ionizing photon counts emitted from galaxies using surveys, together with their impact on the morphology (e.g. size) of their $\HII$ regions (coldspots) using 21 cm tomography. Therefore, this directly probes whether the estimated ionizing photon counts of galaxies and QSOs are sufficient to create the observed size of the $\HII$ region harbouring the sources. This strategy is arguably the most direct test of the sources of reionization.

An alternative strategy, which could already be realized with current observational facilities, is to perform a galaxy survey in regions around high redshift QSOs: while the properties of the ionizing sources would be directly studied using the photometric/spectroscopic data from the surveys, the physical state of the IGM would be inferred from Ly$\alpha$ absorption in the QSO spectra. Observing the ionizing photon production rate and the escape fractions of high-redshift galaxies and QSOs alone does not immediately imply that they are drivers of reionization, as they may be shining in a region already reionized by other sources. As demonstrated in our work, investigating the impact of galaxies and QSOs (as traced by a galaxy survey) on the ionization and thermal states of the IGM (as traced by e.g. QSO near-zone and Ly$\alpha$ absorption line widths) would provide important complementary information about how the sources have driven reionization. For example, \citet{2012MNRAS.423..558C} used the temperature measurement in the local environment of $z\sim6$ QSOs by \citet{2010MNRAS.406..612B} to derive constraints on the ionizing sources.   
In this sense, the data obtained by this strategy (galaxy survey in QSO fields) is best exploited once the interpretation is aided by a RT simulation suite.

\section{Conclusions}\label{sec:conclusion}

We have presented a detailed analysis of hydrogen and helium reionization scenarios in high-$z$ QSO environments including both galaxies and a central QSO using a suite of cosmological multi-frequency radiative transfer simulations. This allows us to understand the array of physical mechanisms that shapes the ionization and thermal states of the IGM. We find that:
\begin{itemize}
\item if the integrated number of ionizing photons emitted by the surrounding galaxies since the onset of reionization exceeds or is comparable to that emitted by the QSOs during its active phase, the morphology of the $\HII$ regions in galaxies only and galaxies+QSO models is similar. On the other hand, if the ionization budget is dominated by the QSO, distinctive features can be observed in the $\HII$ region.       
\item The hard spectrum of the QSO always leaves a unique signature in the morphology of the $\HeII$ and $\HeIII$ regions.
\item Soft X-ray photons from the QSOs produce extended partially ionized tails of hydrogen and helium around the central $\HII$ and $\HeII/\HeIII$ regions. Ionization from secondary electrons enhances the contribution of photoionization by X-rays in the partially ionized tail. On the other hand, neither the morhopology of the fully ionized regions nor the thinckness of the I-fronts are sensitive to the X-ray photons. 
\item UV-obscured QSOs can broaden the I-fronts significantly. A noticeable change in the ionization morphology occurs when soft X-ray photons (from the QSO), rather than UV photons, dominate the growth of I-fronts. 
\item The thermal state of the IGM is strongly affected in a complex, non-linear, way by photoionization heating of both hydrogen and helium. A larger amount of total ionizing photons does not necessarily increase the gas temperature. The highest temperature is attained in the region ionized only by the QSO, as the gas temperature is primarily determined by the spectral type of the sources that first ionized the medium.
\item The X-ray photons from the QSO heat the partially ionized gas ahead of the I-fronts to $T\sim10^3\rm~K$. The net X-ray heating is slightly reduced when secondary ionization is included because it adds another channel to deposit the energy into ionization. The X-ray pre-heating is a clean signature of QSOs activity. 
\end{itemize}

In summary, in the environment surrounding a high-$z$ QSO, {\it the physical state of the IGM in terms of hydrogen and helium reionization as well as gas temperature, is determined by a complex and non-linear interplay of the galaxies and the QSO}. This picture emphasizes the importance of correctly modelling hydrogen, helium and temperature evolution with a multi-frequency treatment of the radiative transfer in 3D numerical simulations.
The large suite of multi-frequency RT simulations presented here provides a highly valuable theoretical resource to understand the physical mechanisms responsible for shaping the ionization and thermal structures of the IGM. This will facilitate the use of RT simulations in interpreting QSO spectra and upcoming 21 cm observations to better understand the role of galaxies and QSOs in driving reionization.

\section*{acknowledgments}

We thank James Bolton for kindly providing us the hydrodynamical simulation which is used in this paper.
KK thanks Michele Sasdelli, Garrelt Mellema, and Ilian Iliev for useful conversations, and Andrew Chung and Phillip Busch for helping to improve the manuscript. LG acknowledges the support of the DFG Priority Program 1573 and from the European Research Council under the European Union's Seventh Framework Programme (FP/2007-2013) / ERC Grant Agreement n. 306476.


\bibliographystyle{mn2e}
\bibliography{Reference}

\appendix

\section{Convergence test}\label{app:photon_packet}

\begin{figure}
 \begin{center}
  \includegraphics[angle=0,width=0.8\columnwidth]{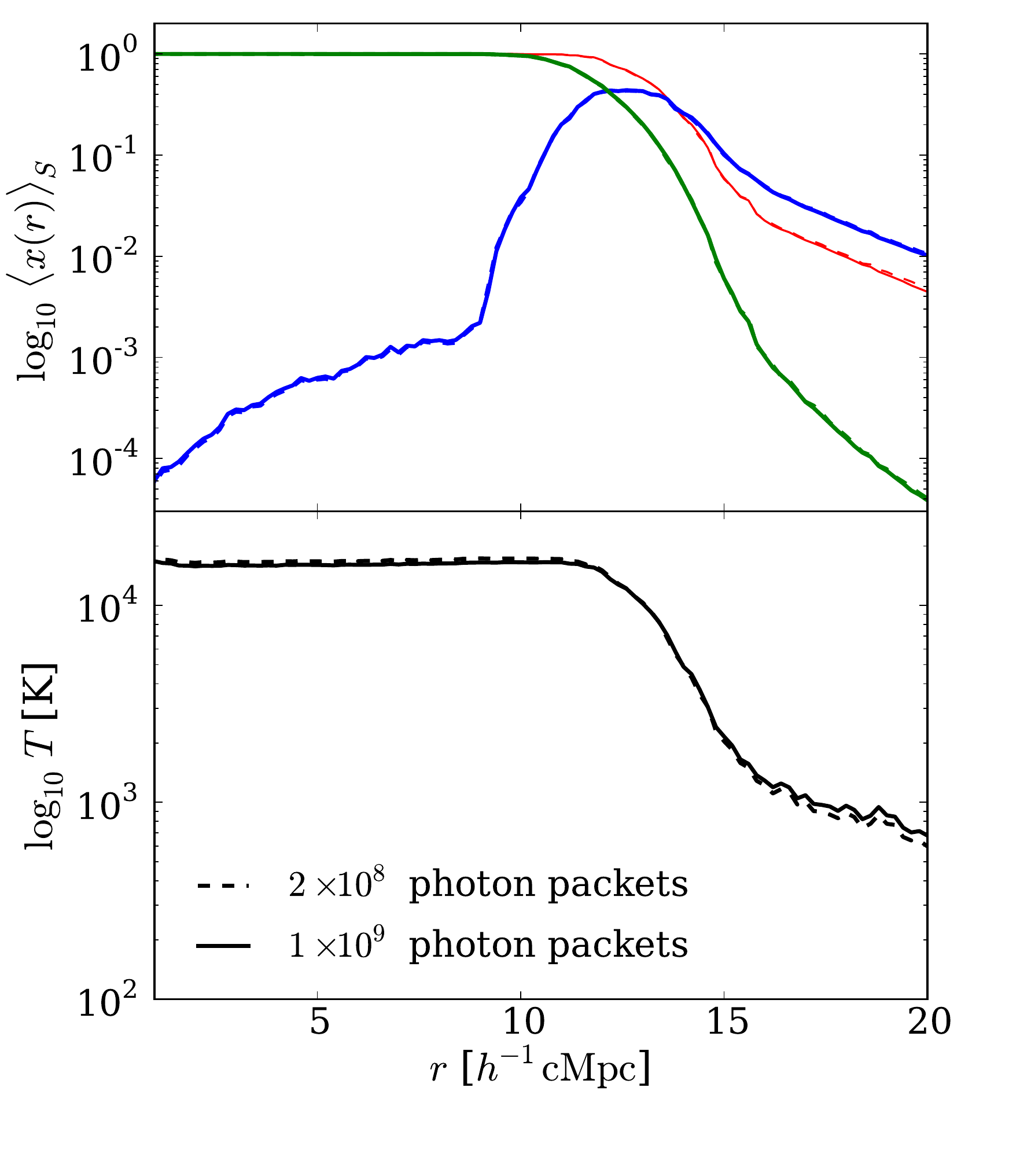}
  \caption{Spherically averaged profiles of $\HII$ (red lines), $\HeII$ (blue), and $\HeIII$ (green) fractions (top panel), and of the temperature (bottom panel) for model QSO\_UVXsec. The line style refers to $2 \times 10^8$ (dashed lines) and $10^9$ (solid) photon packets emitted by the central QSO.}
\label{fig:convergence_test_packets1}
 \end{center}
\end{figure}

\begin{figure}
 \begin{center}
  \includegraphics[angle=0,width=0.8\columnwidth]{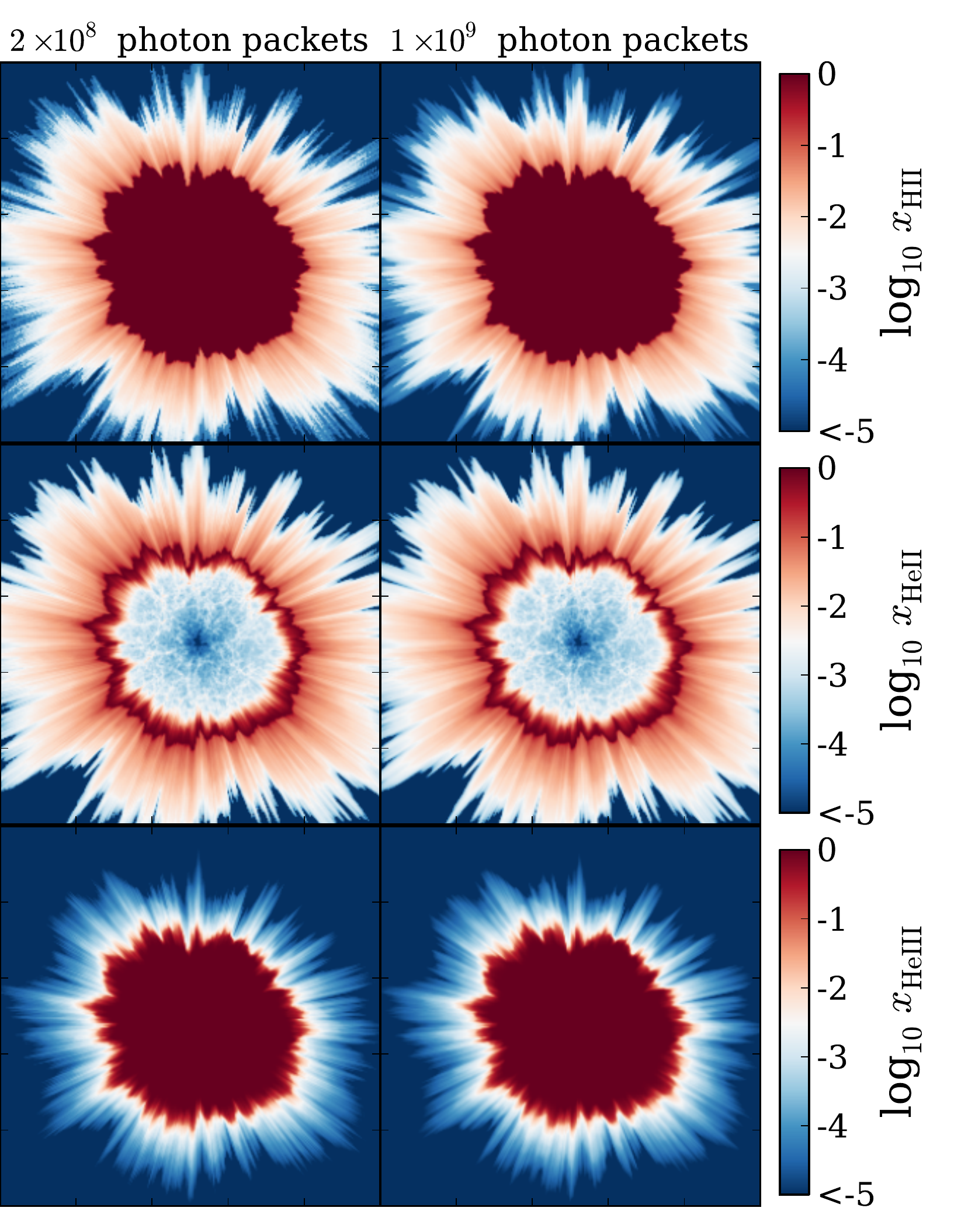}
  \caption{Maps of $\HII$ (top panels), $\HeII$ (middle), and $\HeIII$ (bottom) fractions for model QSO\_UVXsec, where the number of photon packets emitted by the central QSO is $2\times10^8$ (left panels) and $10^9$ (right). The total length of the x- and y-axes is $50h^{-1}\rm cMpc$.}
\label{fig:convergence_test_packets2}
 \end{center}
\end{figure}

We have tested the convergence of our results by increasing the number of photon packets emitted by the QSO from $2\times10^8$ to $10^9$.  The spherically averaged profiles of the ionized fractions and temperature are plotted in Figure \ref{fig:convergence_test_packets1}, and show that, while the number of photon packets is increased by a factor of 5, the resulting profiles are the same for the ionized fractions and differ for the temperature by $5\%$ at most in the partially ionized tail. 
In Figure \ref{fig:convergence_test_packets2} maps of $\HII$, $\HeII$, and $\HeIII$ fractions are shown for the two cases. Although increasing the number of photon packets emitted produces a sligthly smoother tail at a distance of $\sim20h^{-1}\rm cMpc$ from the centre, it is clear that the main fluctuations are not produced by a poor convergence of the Monte Carlo method but rather by shadowing of dense clumps. We thus use $2\times10^8$ photon packets as default value for all our simulations. 

\section{Effect of QSO duty cycle}\label{app:duty}

\begin{figure}
 \begin{center}
  \includegraphics[angle=0,width=\columnwidth]{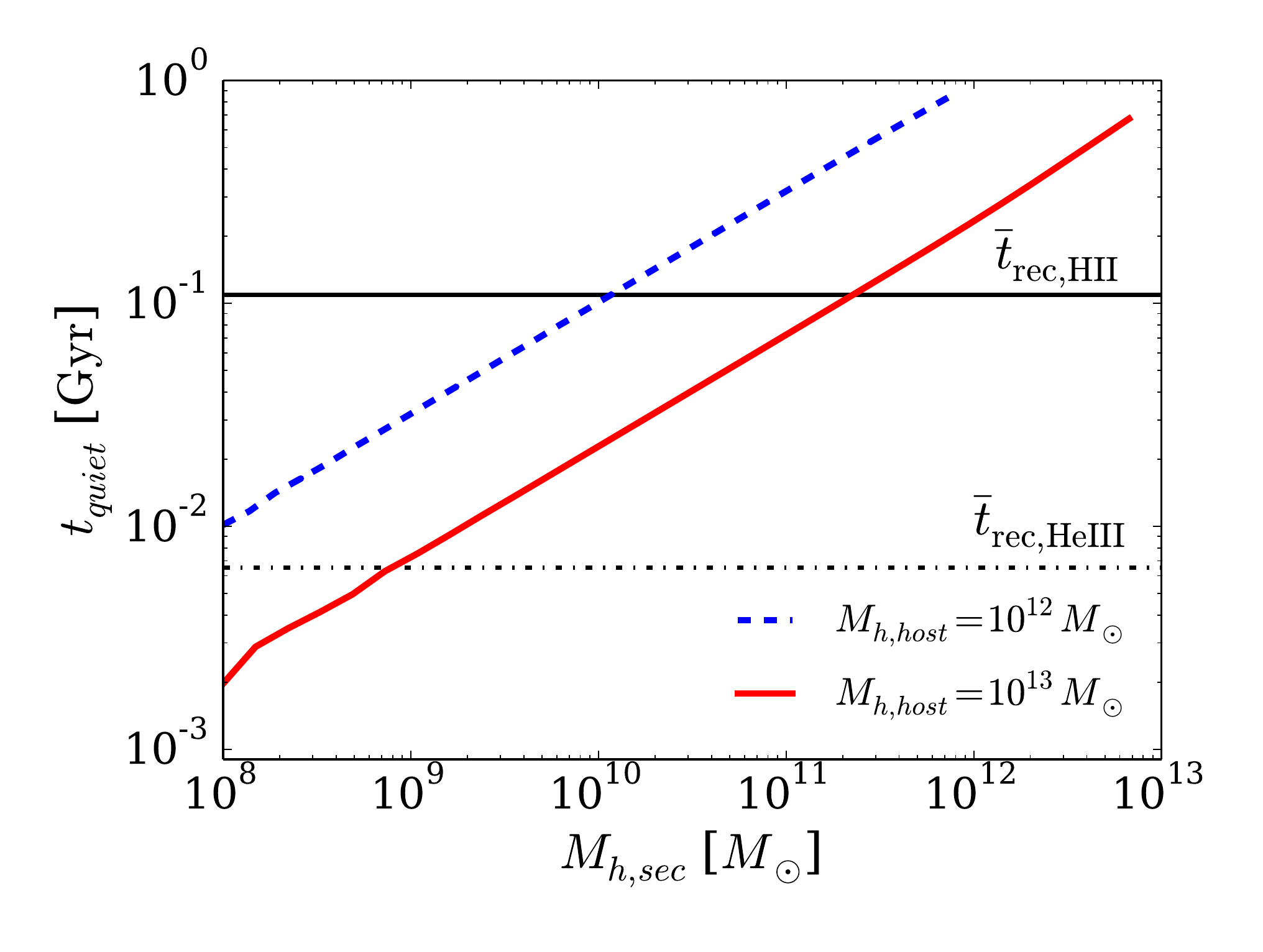}
  \caption[Halo merger timescale and recombination timescale]{Halo merger timescale and recombination timescale as a function of the mass of the secondary halo, $M_{h,sec}$, merging with the halo hosting a QSO. The latter has a mass $M_{h,host}=10^{12}~M_\odot$ (dashed line) or $10^{13}~M_\odot$ (solid). The horizontal solid and dotted-dash lines are the recombination timescale of hydrogen and helium obtained for a clumping factor $C=3$. A different value of $C$ results in a linear upward/downward shift of the horizontal lines.} 
\label{fig:merger_timescale}
 \end{center}
\end{figure}

We investigate the impact of the QSO duty cycle on reionization using an analytic argument, as it allows to scan a large parameter space. Although we have run a simulation including previous activity of the QSO, because the mechanisms triggering it are not fully understood at high redshift, we defer to future work a more extensive numerical investigation. In this argument we exclude the effect of galaxies.

The build-up of a black hole driving the QSO activity requires $t=t_{\rm Edd}\ln(M_{\rm BH}/M_{seed})\sim5\times10^7\rm~yr$, with a seed mass of $M_{seed}=10^{2-5}\rm~M_\odot$ and $t_{\rm Edd}=4\times10^7\rm~yr$ (e.g. \citealt{2013fgu..book.....L}). A QSO-host galaxy may thus have experienced multiple episodes of QSO activity (our fiducial QSO lifetime $t_Q=10^7\rm~yr$ is smaller than the timescale to build up a SMBH). 

As a first approximation, if the duration of the QSO-quiet phase, $t_{quiet}$, is shorter than the recombination timescale of the $i$-th species $i=\HII,\HeIII$ at mean density, i.e. $t_{quiet}<\bar{t}_{rec,i}$, the IGM remains ionized (fossil $\HII$ and $\HeIII$ regions) during the QSO-quiet phase (\citealt{2008ApJ...686...25F,2008MNRAS.384.1080T}). In this case, during the next active phase the QSO turns on in a fossil ionized region, which should thus be taken into account. On the other hand, if $t_{quiet}>\bar{t}_{rec,i}$, the IGM recombines and becomes neutral before the next QSO phase, then the QSO  must ionize again a fully neutral IGM. In this limit, only the phase of QSO activity at the redshift of interest contributes to reionization.

 Assuming that the QSO activity is triggered by merger events of dark matter haloes (e.g. \citealt{2006ApJS..163....1H}), $t_{quiet}$ is expected to be of order of the halo merger timescale. This can be estimated from the extended Press-Schechter approach (\citealt{1993MNRAS.262..627L}) as the inverse of the halo merger rate at $z=10$. Figure~\ref{fig:merger_timescale} shows the QSO-quiet phase timescale for a host with halo mass $M_{h,host}$ merging with a secondary halo of mass $M_{h,sec}$, together with the recombination timescales\footnote{The recombination timescales are calculated as $\bar{t}_{\rm{rec,HII}}=[\alpha_{\rm B,HII}\bar{n}_{\mbox{\tiny H}}(z) C]^{-1}$ and $\bar{t}_{\rm{rec,HeIII}}=[\alpha_{\rm {B,HeIII}}\bar{n}_{\mbox{\tiny He}}(z)  C]^{-1}$, where $C$ is the clumping factor.}.
The figure suggests that to keep the hydrogen ionized (i.e. $t_{quiet}<\bar{t}_{rec,\rm HII}$) the QSO needs to be fed by minor mergers with halo of mass $M_{h,sec}\lesssim10^{10}(10^{11})\rm~M_\odot$ for $M_{h,host}=10^{12}(10^{13})\rm~M_\odot$; otherwise, the QSO $\HII$ region can recombine before the next phase of QSO activity. Because of the shorter recombination timescale of helium, a QSO should be fed even more vigorously to maintain a $\HeIII$ region fully ionized (\citealt{2012MNRAS.419.2880B}). 

In summary, while in general the duty cycle of a QSO affects the ionization of the surrounding medium, its impact is controlled by the ratio between the timescale of the QSO activity and the recombination timescale. QSO-driven $\HII$ regions at $z>6$ can survive when the QSOs are fuelled by frequent minor mergers. On the other hand, if the QSOs are fuelled by less frequent major mergers, the ionized gas has enough time to recombine before the next activity phase, during which the QSO needs to carve a new $\HII$ region. In this case, the contribution of QSOs to reionization is less significant. Because of the shorter recombination timescales of helium, the continous fuelling of QSOs must occur even more vigorously to sustain full helium reionization.

\section{Order of $\HII$, $\HeII$ and $\HeIII$ I-fronts}\label{sec:I-front_order}

The relative position of $\HII$ and $\HeII/\HeIII$ I-fronts essentially determines the morphology and timing of hydrogen and helium reionizations. Interestingly, it depends on the type of ionizing sources, and for a very hard spectrum the I-front of helium can even precede the $\HI$ I-front (\citealt{2007MNRAS.380.1369T}).  

This is well illustrated with the classical equation of cosmological I-front (e.g. \citealt{1987ApJ...321L.107S} see also \S~\ref{sec:clustering}). The relative position of $\HII$ and $\HeIII$ I-fronts of a QSO is
\begin{equation}
\frac{R_{\rm I,QSO}^{\rm HeIII}}{R_{\rm I,QSO}^{\rm HII}}=\left(\frac{X}{Y}\right)^{1/3}\left(\frac{\nu_{\rm HI}}{\nu_{\rm HeII}}\right)^{\alpha_Q/3}.
\end{equation} 
This argument suggests that the $\HeIII$ I-front may be ahead of the $\HI$ I-front if the spectral index is $\alpha_Q<1.8$. It is however an oversimplification and in fact it does not work in our simulations. The analysis in \S~\ref{sec:obscQSO} suggests that a peculiar case in which the $\HeII$ I-front proceeds ahead of the $\HII$ I-front requires an even harder spectrum. Although the $\HeIII$ I-front in the obscured QSO model follows more closely the $\HII$ I-front, we does not see any obvious sign of the inversion (but see \citealt{2008MNRAS.384.1080T}). The order of $\HII$, $\HeII/\HeIII$ I-fronts is an interesting, but not fully investigated topic. Note that this can have an impact on the thermal structure of the IGM because of the different photoionization heating across each I-front.  Thus, this issue should be more thoroughly studied to fully understand the concerted hydrogen and helium reionization scenario in the QSO environment.

\label{lastpage}

\end{document}